\newif\ifsingle
\singlefalse 

\newif\ifarxiv
\arxivtrue

\ifarxiv
	\documentclass[10pt,twocolumn,journal]{IEEEtran}
\else
	\ifsingle
		\documentclass[12pt,onecolumn,journal]{IEEEtran}
	\else	
		\documentclass[10pt,twocolumn,journal]{IEEEtran}
		
	\fi
\fi 

\ifarxiv
	
	\newcommand{\rev}[1]{{\color{black} #1}}
\else
	
	\newcommand{\rev}[1]{{\color{blue} #1}}
\fi

\ifsingle
	 
\else 
	
\fi

\ifsingle
	\def \picscale {1} 
\else 
	\def \picscale {0.4}
\fi

\newif\ifnoappc
\noappctrue

\usepackage[top=1in, bottom=0.75in, left=0.75in, right=0.75in]{geometry}
\usepackage{graphicx}
\usepackage{amssymb}
\usepackage{epsfig}
\usepackage{subcaption} 
\usepackage{epstopdf}
\usepackage{algorithm}
\usepackage{algorithmic}
\usepackage{amsmath}
\usepackage{amsfonts}
\usepackage{dsfont}
\usepackage{url}
\usepackage{chemarr}
\usepackage{chemarrow}
\usepackage{bbold}
\usepackage[version=3]{mhchem}
\usepackage{mathabx}
\usepackage[dvipsnames]{xcolor}
\usepackage{comment}

\graphicspath{{./z_pic}{./}{./z_pic_arxiv}}

\begin{document} 

\title{Enzymatic cycle-based receivers for approximate maximum a posteriori demodulation of concentration modulated signals} 

\ifarxiv
\author{Chun Tung Chou \\
School of Computer Science and Engineering, University of New South Wales, Sydney, New South Wales 2052, Australia. \\
E-mail: c.t.chou@unsw.edu.au}

\fi

\maketitle

\begin{abstract}
Molecular communication is a bio-inspired communication paradigm where molecules are used as the information carrier. This paper considers a molecular communication network where the transmitter uses concentration modulated signals for communication. Our focus is to design receivers that can demodulate these signals. We want the receivers to use enzymatic cycles as their building blocks and can work approximately as a maximum a posteriori (MAP) demodulator. No receivers with all these features exist in the current molecular communication literature. We consider enzymatic cycles because they are a very common class of chemical reactions that are found in living cells. In addition, a MAP receiver has good statistical performance. 
\rev{
In this paper, we study the operating regime of an enzymatic cycle and how the parameters of the enzymatic cycles can be chosen so that the receiver can approximately implement a MAP demodulator. We use simulation to study the performance of this receiver. We show that we can reduce the bit-error ratio of the demodulator if the enzymatic cycle operates in specific parameter regimes.}  
\end{abstract}


\section{Introduction}
\label{sec:intro} 
Molecular communication is a bio-inspired communication paradigm where the transmitters and receivers use molecules to communicate with each other \cite{Akyildiz:2008vt,Nakano:2014fq,Farsad:2016eua}. One can take this bio-inspiration a step further by considering the fact that living cells encode and decode molecular signals by using molecular circuits, or sets of chemical reactions. This has motivated researchers in molecular communications to study and design chemical reaction-based transmitters and receivers \cite{Bi:CST:2021,Jamali:ProcIEEE:2019,Femminella:DSP:2022}. This paper focuses on designing a reaction-based demodulator for molecular communications. 

There is a growing list of work in molecular communications that uses reaction-based receivers. Kuscu and Akan \cite{Kuscu:TCom:2019} designed a molecular circuit that can extract information from multiple types of ligand. We designed in \cite{Chou:2019gf} a molecular circuit which can approximately perform maximum a posteriori (MAP) demodulation and we shown later on in \cite{Chou:PRE:2022} that the circuit can be implemented by gene promoters with multiple binding sites. Bi et al.~\cite{Bi:TCom:2020} designed a molecular receiver which uses a catalytic-like reaction to amplify the received molecular signal and the concept was later implemented in a microfluidic testbed in \cite{Walter:NatureComm:2023}. Heinlein et al.~\cite{Heinlein:NanoCom:2023} derived a reaction-based realisation of the MAP demodulator by exploiting a connection between MAP and Boltzmann machines. A gap in the existing literature on reaction-based receiver design is that none of the designs is based on enzymatic cycles (e.g. phosphorylation-dephosphorylation cycles, methylation-demethylation cycles) which are a very common class of chemical reactions in the living cells \cite{Alberts}. In addition, synthetic biologists have started to build synthetic protein circuits \cite{Gao.2018kp}. A goal of this paper is to study how receivers based on enzymatic cycles can be designed. 

\begin{figure}
\begin{center}
\includegraphics[trim=0cm 0cm 0cm 0cm, clip=true, width=\columnwidth]{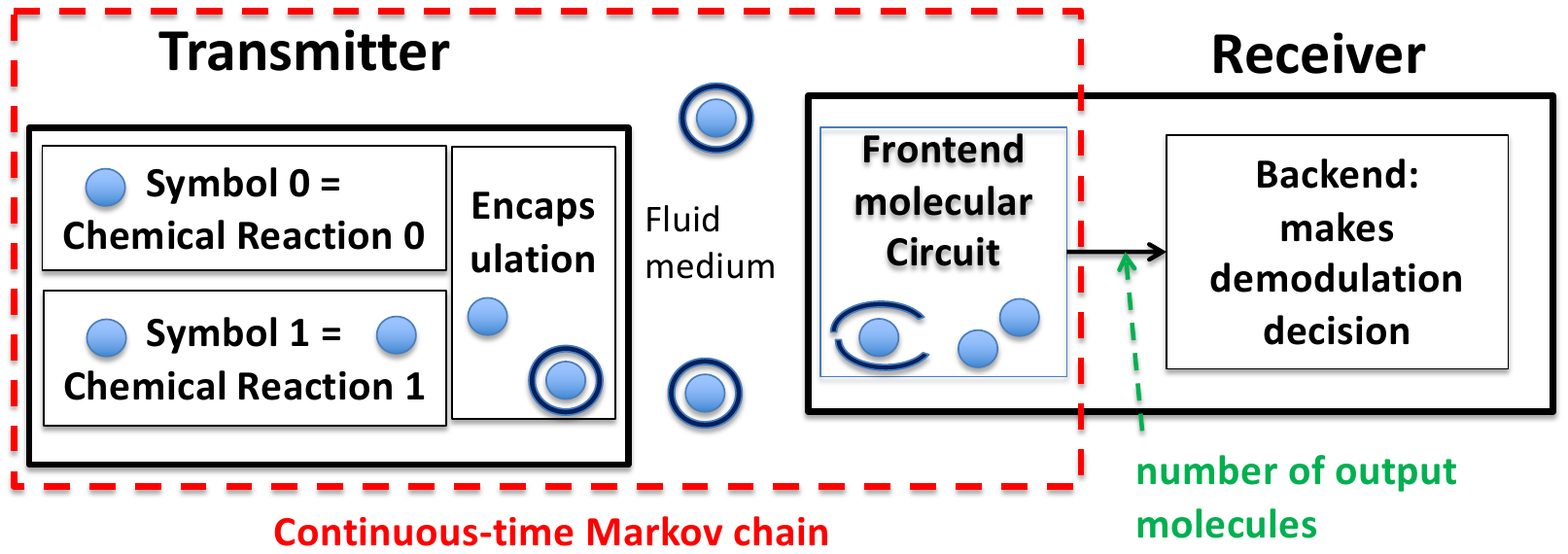}
\caption{Overview of the communication elements (transmitter, receiver and medium) and the signalling molecules. Light blue filled circles depict signalling molecules. A filled circle with a dark blue ring depicts a signalling molecule within a vesicle. The broken ring in the receiver depicts the release of the signalling molecule from a vesicle. }
\label{fig:system_overview}
\end{center}
\end{figure}

This paper is built upon the framework in our earlier work \cite{Chou:2015ga,Awan:2017fm} which uses a Markovian approach to design MAP demodulators. The work \cite{Awan:2017fm} assumes that the receiver consists of two blocks in series: a front-end and a back-end, as in Fig.~\ref{fig:system_overview}. The front-end is a molecular circuit which reacts with the signalling molecules from the transmitter to produce output molecules. The back-end works as a MAP demodulator by using the number of output molecules over time to compute the the log-posteriori probabilities of the possible transmission symbols. The contribution of \cite{Awan:2017fm} is to derive an ordinary differential equation (ODE) which governs the time-evolution of the log-posteriori probabilities given the front-end molecular circuit. Note that the results in \cite{Awan:2017fm} are very general as the front-end can be any set of chemical reactions. Since our goal is to design a receiver that uses enzymatic cycles, we will choose the front-end to be an enzymatic cycle.  
\rev{As the result in \cite{Awan:2017fm} derives the MAP modulator for a given set of front-end circuit parameters, therefore we need to choose good front-end parameters in order to achieve overall optimality of the receiver. However, this is a hard optimization problem because an intermediate step to obtain the MAP modulator requires a Bayesian filtering problem to be solved and this problem does not have a closed-form solution \cite{Bronstein:2018eh}. In this paper, we propose a simplified approach. We propose to use the front-end parameters that enhance its sensitivity to the transmission symbols. Given this front-end design, we then use \cite{Awan:2017fm} to derive the ODE for computing the log-posteriori probabilities. We then show how we can use enzymatic cycles to approximately realise this ODE. We show that the proposed method improves the bit error rate (BER) of the receiver. 
}  
This paper makes the following contributions:
\begin{itemize} 
\item \rev{We explore the parameter space of a front-end enzymatic circuit and show how we can choose its parameter to improve its sensitivity to the transmission symbols.} 
\item We derive a method to approximately compute the log-posteriori probabilities which will later on lead to an implementation using enzymatic cycles. The approximation consists of multiple steps. In one of these steps, we derive a closed-form approximation of an optimal Bayesian filtering problem. \rev{In another step, we show how we can approximate the log-posteriori probability computation.} 
\item We design an enzymatic circuit, which is composed of three enzymatic cycles, that can approximately compute the ratio of log-posteriori probabilities. We demonstrate the accuracy of this approximation using simulation. 
\item \rev{We show that the front-end circuits with higher sensitivity give an overall MAP receiver that have a better BER.}
\end{itemize}

The rest of the paper is organised as follows. Sec.~\ref{sec:related} discusses related work. In Sec.~\ref{sec:bg}, we present the set up of our molecular communications problem and relevant background results from \cite{Awan:2017fm}. Sec.~\ref{sec:pdp:approx} present results on 
\rev{sensitivity improvement}, log-posteriori probability approximation and the design of the enzymatic circuit. We then present simulation results in Sec.~\ref{sec:eval} and conclude in Sec.~\ref{sec:con}. 




\section{Related Work} 
\label{sec:related} 
\rev{
This paper falls within a larger theme of using chemical reactions in molecular communications. There has been an increase in activities on using chemical reactions in molecular communications and we refer the reader to a few surveys \cite{Bi:CST:2021,Jamali:ProcIEEE:2019,Femminella:DSP:2022} which focus on this line of research. There are multiple ways that chemical reactions have been used in molecular communications. We can classify them into three categories, depending on whether the focus is on the transmitter, medium or receiver. On the transmitter side, \cite{Deng:2017uw} uses chemical reactions to produce transmission signals for molecular communication and \cite{Arjmandi:TNB:2016} considers a transmitter that uses ion channels. In addition, communication performance can also be improved by using chemical reactions in the channel \cite{Farahnak:TCOM:2019}. Although the above works  use chemical reactions at the transmitter or in the medium, the focus of our work is on the receiver side. 

The focus of this paper is on designing chemical reaction-based receivers. We have already discussed a number of representative work \cite{Kuscu:TCom:2019}\cite{Bi:TCom:2020}\cite{Walter:NatureComm:2023}\cite{Heinlein:NanoCom:2023} in the introduction. Our work differs from the existing work in two major aspects.  First, the earlier work assumed that the demodulation is based on one sample point per symbol; however, this work assumes that demodulation is based on the continuous history of the number of active receptors. We showed in \cite{Chou:2019gf} that demodulation using a continuous history gives a lower BER in comparison. Second, the molecular demodulator considered in earlier work did not use enzymatic cycles. To the best of our knowledge, there is few work  \cite{Awan.2018} \cite{Ratti:Molecules:2022} \cite{McBride:PIEEE:2019} on considering the use of enzymatic cycles in molecular communication. The work \cite{Awan.2018} and \cite{Ratti:Molecules:2022} focused on channel capacity, rather than the design of enzymatic circuits; and \cite{McBride:PIEEE:2019} focuses on how an upstream enzymatic circuit can affect the performance of a downstream circuit. Therefore, in comparison to the state of the art in chemical reaction-based receivers, an advance made by this paper is to design enzymatic circuits that can act as communication receivers.
}


In the molecular communication literature, molecular circuits have also been studied for various applications. For example, \cite{MrAlessioMarcone:2017te,Marcone:2018kp} present genetic circuits for parity-check. As another example, \cite{Lombardo:TMBMC:2024,Veletic:TNB:2020} studied molecular circuits for inter-cellular communication and targeted drug delivery.\rev{The key difference between these work and our is that we use stochastic, rather than deterministic, analysis.}

\section{Model and Exact Computation of Posteriori Probabilities} 
\label{sec:def}
\label{sec:bg} 

This section presents the set up of the following three components in our molecular communication system: the medium, the transmitter and the front-end of the receiver. Note in particular, the front-end of the receiver is an enzymatic cycle. By using this set up and our earlier work \cite{Awan:2017fm}, we present an ODE which describes the evolution of posteriori probability over time. This ODE forms the basis of this paper and our goal is to show how we can realise this ODE by using enzymatic cycles in Sec.~\ref{sec:pdp:approx}. 

\subsection{Transmission Medium and Transmitter}
\label{sec:model:medium_and_TX}
The modelling framework of this paper mostly follows our previous work \cite{Chou:2015ga}\cite{Chou:gc}. We model the medium as a rectangular prism and divide the medium into voxels. We assume that the transmitter and the receiver each occupies a voxel. Note that it is possible to generalise to the case where a transmitter or receiver consists of multiple voxels, see \cite{Riaz:TCom:2020}, but we have not done that to simplify the presentation. Although it may not be physically realistic for the receiver to have a cubic shape, this simplified geometry allows us to focus on the signal processing aspect of the receiver. 

We assume the transmitter communicates with the receiver using one type of signalling molecule \cee{K}, which is depicted as light blue circles in Fig.~\ref{fig:system_overview}. \rev{With the use of enzymatic circuits at the receiver, the signalling molecule is an enzyme. However, the size and physical properties (e.g., polarity) of some enzymes may not allow them to pass through the membrane of cells unaided. We therefore assume that the transmitter encapsulates these signalling molecules \ce{K} in vesicles (denoted as ${\hat{\rm K}}$) for transportation and to help their entrance into the receiver, e.g., via endocytosis. In Fig.~\ref{fig:system_overview}, we use a dark blue ring to depict the vesicle that encapsulates the signalling molecule. We remark that researchers have studied the use of artificially made extra-cellular vesicles to transport various biological molecules including protein \cite{Ziegler:EV:2023} and the idea in this paper fits into this context. } 

\rev{
We assume that the vesicles randomly diffuse in the medium with a diffusion coefficient $D$. If a voxel is a cube with a volume of $W^3$, then the infinitesimal probability that a vesicle diffuses to one of the six neighbouring voxels is $\frac{D}{W^2} (\Delta t)$ where $\Delta t$ is an infinitesimal time. Note that this probability only holds if the neighbouring voxel is not a receiver voxel. We will discuss the transport of the vesicle into the receiver in Sec.~\ref{sec:model:rx_front_end}.} 

We assume that the transmitter uses $S = 2$ different symbols indexed by $s = 0, 1$. (Generalisation to the $S > 2$ case is left for further research.)
\rev{Furthermore, we assume the signalling molecules for these symbols are produced by some chemical reactions in the transmitter (see Fig.~\ref{fig:system_overview}) and each signalling molecule is encapsulated in a vesicle before it is released into the medium. The release rate $\rho_s$ of the vesicles into the medium is assumed to be a constant for each symbol $s = 0, 1$ and we assume $\rho_1 > \rho_0$. In other words, the transmitter uses concentration shift keying. 
} 


\subsection{Receiver: modelling and its front-end molecular circuit} 
\label{sec:model:rx_front_end} 
\rev{
If a vesicle hits the surface of the receiver voxel, there is a probability that it enters the receiver using endocytosis. We assume that if a vesicle successfully enters the receiver, then the signalling molecule will be released from the vesicle as illustrated by the broken vesicle shell in Fig.~\ref{fig:system_overview}. These freed signalling molecule \ce{K} (depicted as the blue filled circles in the receiver in Fig.~\ref{fig:system_overview}) can now react with the front-end receiver molecular circuit which we will describe shortly. The molecule \ce{K} will stay in the receiver until it is degraded. 
}

\rev{
If a vesicle is in a neighbouring voxel of the receiver voxel, we assume that there is an infinitesimal probability of $\chi \frac{D}{W^2} (\Delta t)$ that this vesicle entering the receiver voxel. We will use a small value of $\chi$ to reflect the fact that it is harder for a vesicle to enter the receiver voxel than to diffuse to a neighbouring voxel. 
}

The receiver front-end is assumed to be an enzymatic cycle which reacts with the signalling molecule \ce{K}. \rev{This cycle reacts only with \ce{K} in its freed form and does not react with the vesicle that carries \ce{K}.}  The cycle consists of three species \ce{X}, \ce{XK} and \ce{X_*} which take part in the following four reactions: 
\begin{subequations}
\label{cr:p:z0:all} 
\begin{align}
\cee{
X + K &  <=>[a_0][d_0] XK \label{cr:p:z0:1}  \\
XK &  ->[g_0] X_* + K  \label{cr:p:z0:2} \\
X_* &  ->[g_-]   X  \label{cr:p:z0:3}  
} 
\end{align}
\end{subequations}
where $a_0$, $d_0$, $g_0$ and $g_-$ are reaction rate constants. In the case where the above reactions are of the phosphorylation-dephosphorylation reaction type, then we identify \ce{X}, \ce{K}, \ce{XK} and \ce{X_*} as, respectively, a unphosphorylated substrate, kinase, complex and phosphorylated substrate \cite{Gomez-Uribe:PLoS_CB:2005}. We assume that the \textsl{substrate} --- which is the collection of all the species that has an \cee{X} in them, i.e., \cee{X}, \cee{XK} and \cee{X_*} --- can only be found in the receiver and cannot be diffused outside of the receiver. 


Since the transmitter emits \rev{vesicles} at different rates for Bits 0 and 1, this will create different concentration levels of \cee{K} in the receiver voxel. We would like the reactions in \eqref{cr:p:z0:all} to produce a low (resp. high) number of \cee{X_*} when Bit 0 (Bit 1) is sent to enable the receiver to infer the bit that the transmitter has sent by using the amount of \ce{X_*}. In fact, \rev{one may consider the the purpose of the front-end is to map the concentration of \ce{K} to a concentration of \ce{X_*} via the reactions above. After that, the back-end will use the amount of \ce{X_*} to infer which symbol has been sent.} We want this \rev{map} to be sensitive to the amount of \cee{K} \rev{in the sense that the concentration difference of \ce{K} due to the two transmission symbols will result in maximum difference in concentrations of \ce{X_*}.} We will show in  Sec.~\ref{sec:pdp:approx:front_end:para} how we can choose the rate constants in \eqref{cr:p:z0:all} to achieve high sensitivity. 






We have now described the transmitter, the medium and the receiver front-end. These system components have reactions (e.g., the reactions \eqref{cr:p:z0:all} in the receiver voxel, \rev{the entrance of a vesicle into the receiver as an equivalent first order reaction}) and diffusion of the vesicles. We model the \rev{stochastic} dynamics of these three components (which is the dashed box in Fig.~\ref{fig:system_overview}), \rev{which include both diffusion and reactions}, by using the reaction-diffusion master equation (RDME)\footnote{There are three major classes of stochastic models for modelling systems with both diffusion and reactions. They are the Smoluchowski equation, RDME and the Langevin equation \cite{Erban:2007we}. The Smoluchowski equation is based on particle dynamics. It is a fine grained model but hard to work with analytically. Both RDME and Langevin are easier to work with analytically but master equation has a finer scale and granularity compared to the Langevin equation \cite{DelVecchio:book}. Therefore we choose to use RDME which allows us to use Markovian theory for analysis and is at the same time a finer grained model.} \cite{Gardiner} (which is a specific type of continuous-time Markov chain). The receiver back-end is used for MAP demodulation, which will be discussed in Sec.~\ref{sec:model:MAP_demodulation}. 

\subsection{Notation} 
We will deal with time-varying chemical signals in this paper. For a chemical species, we will use its species name in \textsl{italics} font to denote its count. For example, the count of \cee{X_*} molecules in the receiver at time $t$ is denoted by $X_*(t)$. This applies to all chemical species with one letter. For chemical species with multiple letters, we add curly brackets $\{ \; \}$ around their name, e.g. $\{XK\}(t)$ is the molecular count of the species \ce{XK} at time $t$. 

We will use steady state analysis \cite{DelVecchio:book} to help us to understand the properties of the receiver. Typically steady analysis expresses its results in concentration. We will denote the concentration of a species by enclosing its name within a pair of square brackets $[\;]$, e.g. $[X_*]$ denotes the concentration of \ce{X_*}. For the RDME model, both molecular counts and concentration are studied at the spatial scale of a voxel and we can convert between them using the volume of a voxel $\Omega$, e.g. $[X_*] = \frac{X_*}{\Omega}$.  

We add the superscript $^{\rm ss}$ to a quantity to denote its steady state mean value, e.g. $X_*^{\rm ss}$ is the mean number of \ce{X_*} molecules at steady state. 

Sometimes we may need to express the molecular count of a particular species due to a specific transmission symbol $s$ where $s \in \{0,1\}$. In that case, we add $s$, $0$ or $1$ as a subscript to indicate that, e.g., $X_{*,1}^{\rm ss}$ is the steady state count of \ce{X_*} when symbol 1 is transmitted. 

\rev{
The vesicle $\hat{\rm K}$ diffuses in the medium so each voxel has its own molecular count of $\hat{\rm K}$. Although the signalling molecules \ce{K} are found in both transmitter and receiver, the results will only require us to consider the count or concentration of \ce{K} in the receiver voxel.} We will use $K(t)$, $[K]$ to indicate the count and concentration \ce{K} in the receiver voxel. 

Lastly, the total count of some species may be conserved in the receiver voxel. For example, the total count of \ce{X}, \ce{XK} and \ce{X_*} is a constant. We use $X_T$ and $[X]_T$, with a subscript $T$ which is short for ``total", to denote the conserved total count and concentration.  

\subsection{MAP Demodulation: Problem and Solution}
\label{sec:model:MAP_demodulation} 
In this paper, we consider a demodulation problem of using the information on \cee{X_*} to infer the symbol that the transmitter has sent. We will focus on the demodulation of one symbol \rev{by assuming that the signalling molecules \ce{K} from earlier symbols have degraded by the time a new symbol is received.} \rev{We will leave the problem of inter-symbol interference as future work.} 

In the formulation of the demodulation problem, we will assume that at time $t$, the data available to the demodulation problem are $X_*(\tau)$ for all $\tau \in [0,t]$; in other words, the data are continuous in time and are the history of the counts of \cee{X_*} up to time $t$. We will use ${\cal X}_*(t)$ to denote the continuous-time history of $X_*(t)$ up to time $t$. Given that we model the molecular communication system using RDME, this means that $X_*(t)$ is a realisation of a continuous-time Markov chain. 

We adopt a MAP framework for detection. Let ${\mathbf P}[s | {\cal X_*}(t)]$ denote the posteriori probability that symbol $s$ has been sent given the history ${\cal X_*}(t)$. Since we assume the transmitter uses only 2 symbols, the demodulation decision can be made from using the log-probability ratio $L(t)$:
\begin{align}
L(t) & = \log \left( \frac{{\mathbf P}[1 | {\cal X_*}(t)]}{{\mathbf P}[0 | {\cal X_*}(t)]} \right)
\label{eqn:def:L}
\end{align}
If the demodulation decision is to be done at time $t$, then the demodulator decides that Symbol 1 has been sent if $L(t)$ is greater than a pre-defined threshold. By using the method in \cite{Awan:2017fm}, we show in 
\ifarxiv
Appendix \ref{app:sol:log_post_prob} 
\fi
that the evolution of the log-probability ratio $L(t)$ obeys the following ODE: 
\begin{align}
\frac{dL(t)}{dt} 
= & \left[ \frac{ dX_*(t) }{dt} \right]_+
\log \left( \frac{J_1(t_-)}{J_0(t_-)} \right) - k_0  \left( J_1(t) - J_0(t) \right) 
\label{eqn:ode:L}  
\end{align}
where $[w]_+ = \max (w,0)$. The quantity  
$J_s(t) =  {\rm E}[\{XK\}(t) | s, {\cal X}_*(t)]$ is the posteriori mean of the number of \cee{XK} molecules at time $t$ given the history ${\cal X}_*(t)$ and the assumption that the transmitter has sent Symbol $s$. We can determine $J_s(t)$ by solving an optimal Bayesian filtering problem \cite{Sarkka}. We assume that the two transmission symbols are equally likely so $L(0) = 0$. 

The MAP demodulator is located in the back-end of the receiver in  Fig.~\ref{fig:system_overview}. In the next section, we will show that it is possible to use enzymatic cycles to approximately realize the ODE in \eqref{eqn:ode:L} and hence the back-end. 


\section{Designing an enzymatic cycle-based receiver} 
\label{sec:pdp:approx}
In this section, we use enzymatic cycles as the circuit components to realise a receiver that can approximately compute the log-posteriori probability in ODE \eqref{eqn:ode:L}. The overall receiver design consists of three enzymatic cycles. One cycle is used in the front-end while the other two are used in the back-end. 
\rev{Our design starts in Sec.~\ref{sec:pdp:approx:front_end:para} where we study the choice of the receiver front-end parameters to improve its sensitivity. 
}
We next move onto designing the back-end which is to realise \eqref{eqn:ode:L} using enzymatic cycles. This consists of two steps. 
The first step, which is in Sec.~\ref{sec:approx:filtering}, is to derive an approximation of \eqref{eqn:ode:L} which removes the computational demanding parts of \eqref{eqn:ode:L}. We then show in Sec.~\ref{sec:pdp:circuit} how we can use  enzymatic cycles to realise the approximation of \eqref{eqn:ode:L} that we have derived in Sec.~\ref{sec:approx:filtering}. 

\subsection{Parameters of receiver front-end} 
\label{sec:pdp:approx:front_end:para} 
\begin{figure}
\begin{center}
\includegraphics[trim=0cm 0cm 0cm 0cm, clip=true, width=\columnwidth]{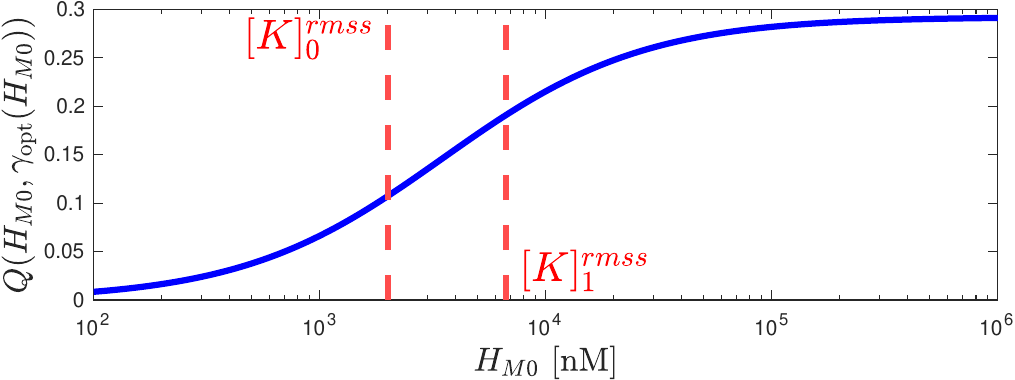}
\caption{\rev{A plot of $Q(H_{M0},\gamma_{\rm opt}(H_{M0}))$. }}
\label{fig:sensitivity:H_M0}
\end{center}
\end{figure}

\rev{
Fig.~\ref{fig:system_overview} shows that the complete receiver consists of a front-end and a back-end where the front-end is the enzymatic cycle \eqref{cr:p:z0:all} and the back-end is the MAP modulator. Note that the MAP demodulator is derived for a specific  set of front-end circuit parameters. Therefore, if we want to achieve overall optimality of the entire receiver, we need to optimise the choice of the front-end parameters. However, this is a formidable task because the log-probability ratio in \eqref{eqn:ode:L} does not exist in closed-form. In this section, we present a heuristic argument to maximise the sensitivity of the front-end receiver and we will verify in Sec.~\ref{sec:eval} that our heuristically chosen front-end parameters produce receivers with better BER. 

Given that the transmitter uses concentration shift keying, the two symbols will result in two different levels of concentration of $K$ at the receiver and we will refer to them as $[K]_s^{\rm ss}$ for $s = 0, 1$. Since we assume \ce{X_*} is the output molecule, this means the MAP estimator uses the counts of \ce{X_*} to distinguish between the two symbols. Thus, if the two symbols result in a very different value of steady state concentration levels $[X_{*}]_s^{\rm ss}$, then there is a better chance of distinguishing between them. We propose to use $[X_{*}]_1^{\rm ss} - [X_{*}]_0^{\rm ss}$ as a sensitivity measure of the front-end. In Appendix \ref{app:front_end}, we use steady state analysis to show that:
\begin{eqnarray} 
[X_*]_s^{\rm ss} & = & [X]_T 
\frac{\gamma [K]_s^{\rm ss}}{H_{M0} + (1 + \gamma) [K]_s^{\rm ss}}
\label{eqn:Xstar:perturbation}
\end{eqnarray}
where $H_{M0} = \frac{d_0 + g_0}{a_0}$ and $\gamma = \frac{g_0}{g_-}$.
(Note that $H_{M0}$ is known as a Michaelis-Menten constant in the chemistry literature, hence the subscript ``M".) This shows that, despite the fact that the enzymatic cycle \eqref{cr:p:z0:all} has 5 reaction constants, the steady state $[X_*]_s^{\rm ss}$ depends on $[K]_s^{\rm ss}$ via two parameters  $H_{M0}$ and $\gamma$. 

Our aim is to choose $H_{M0}$ and $\gamma$ to maximise the sensitivity $
Q(H_{M0},\gamma) \triangleq [X_{*}]_1^{\rm ss} - [X_{*}]_0^{\rm ss}$ assuming that $[K]_s^{\rm ss}$ (for $s = 0, 1$) are given. We first consider maximising the sensitivity $Q$ by using $\gamma$ while holding $H_{M0}$ constant. We can show that, for a given $H_{M0}$, the sensitivity $Q$ is maximised by choosing 
$\gamma$ to be $\gamma_{\rm opt}(H_{M0}) = \frac{1}{\sqrt{\xi_0 \xi_1}}$ where $\xi_s = \frac{[K]_{s}^{\rm ss}}{H_{M0} + [K]_{s}^{\rm ss}}$. It can readily be seen this $\gamma > 1$. With this expression of $\gamma$, we can find the maximum sensitivity for each value of $H_{M0}$ and we have plotted $Q(H_{M0},\gamma_{\rm opt}(H_{M0}))$ in Fig.~\ref{fig:sensitivity:H_M0}. It shows that the sensitivity can be increased by using a large value of $H_{M0}$ but sensitivity does not increase much beyond a certain value of $H_{M0}$. We can show mathematically that $Q(H_{M0},\gamma_{\rm opt}(H_{M0}))$ is an increasing function of $H_{M0}$ and we need $H_{M0} \gg [K]_1^{\rm ss}$ for high sensitivity, see  Fig.~\ref{fig:sensitivity:H_M0}. Note that the typical range of Michaelis-Menten constant is 100-10$^8$ nM \cite{libretextsMichaelisMentenKinetics} so we will use values in this range in our evaluation. 

By using the assumption $H_{M0} \gg [K]_1^{\rm ss}$, we show in Appendix \ref{app:front_end} that the number of \ce{XK} molecules is small. Let us recall that \ce{XK} is a complex formed by the binding of a \ce{K} molecule to an \ce{X} molecule. Once a \ce{K} molecule has been bound to an \ce{X} molecule to form an \ce{XK} molecule, this \ce{K} molecule can no longer be used as it is sequestered with an \ce{X}. Since \ce{XK} is the only molecule in the front-end \eqref{cr:p:z0:all} that sequesters the signalling molecule \ce{K}, so a small \ce{XK} means few \ce{K} molecules are sequestered. 
We will need the requirement of few sequestered \ce{K} molecules in Sec.~\ref{sec:Lhat:X_times_Y} to argue the correctness of our enzymatic circuit realisation of the entire receiver and this requirement can be achieved if the Michaelis-Menten constant $H_{M0}$ is sufficiently large. Therefore, having a sufficiently large $H_{M0}$ is central requirement and we will focus on obtaining a method to approximately compute the log-probability ratio \eqref{eqn:ode:L} in the region where $H_{M0} \gg [K]_1^{\rm ss}$.
} 

Note that the arguments in this section and Appendix \ref{app:front_end} is based on deterministic chemical rate equations. We will use stochastic simulation in Sec.~\ref{sec:eval} to show that the number of \ce{XK} molecules is small for the stochastic case. 

We have now discussed how the parameters of the front-end enzymatic cycle can be chosen. In the next sub-section, we will explain how we can turn \eqref{eqn:ode:L} into a form which can be realised by enzymatic cycles. 

\subsection{Approximating \eqref{eqn:ode:L}} 
\label{sec:approx:filtering}
The ODE \eqref{eqn:ode:L} is not in a form that can be readily realised by enzymatic cycles because it contains complex mathematical operations. In this section we will use a few approximation steps to turn \eqref{eqn:ode:L} into a form which can be realised by an enzymatic circuit in Sec.~\ref{sec:pdp:circuit}. 

In the first step, we replace the posteriori mean $J_s(t) =  {\rm E}[\{XK\}(t) | s, {\cal X}_*(t)]$ in \eqref{eqn:ode:L} by a closed-form approximation. In Appendix \ref{app:filtering}, we show that if $H_{M0} \gg [K]_{1}^{\rm ss}$, which is the same assumption needed for the receiver front-end to have high \rev{sensitivity}, then  $J_s(t)$ can be approximated by:
\rev{
\begin{align}
J_s(t) \approx 
\underbrace{
\frac{[K]_{s}^{\rm ss}}{g_0 \frac{X_T - X_{*,s}^{\rm ss}}{\Omega} +  [K]_{s}^{\rm ss} + H_{M0}} }_{\kappa_s}  
(X_T - X_{*}(t))
\label{eqn:Js:approx} 
\end{align}
}
where $X_{*,s}^{\rm ss}$ is the steady-state mean number of \ce{X_*} molecules when the transmitter sends Symbol $s$. 
Note that $\kappa_s$ depends only on the transmitter, medium and front-end parameters, so it is independent of the measurements $X_*(t)$. 

After substituting \eqref{eqn:Js:approx} into \eqref{eqn:ode:L}, we have: 
\begin{align}
\frac{dL(t)}{dt} 
= & \left[ \frac{ dX_*(t) }{dt} \right]_+
\log \left( \frac{\kappa_1}{\kappa_0} \right) - k_0  \left( \kappa_1 - \kappa_0 \right) (X_T - X_*(t))
\label{eqn:ode:L:kappa}  
\end{align}
The above ODE is still difficult to realise by chemical reactions. In particular, the existing methods to implement derivatives \cite{Alexis:iScience:2021} and subtraction \cite{Oishi:2011ig} (or specifically the subtraction in $(X_T - X_*(t))$) require additional chemical species and reactions to be introduced. In 
\ifarxiv
Appendix \ref{app:Lhat}
\fi
, we show that, if the receiver front-end has high \rev{Michaelis-Menten constant $H_{M0}$}, then we can approximately compute $L(t)$ in \eqref{eqn:ode:L:kappa} by using:  
\begin{align}
\frac{dL(t)}{dt} \approx & 
g_- X_*(t) 
\left[
\log \left( \frac{\kappa_1}{\kappa_0} \right) - H_{M0} \Omega \frac{ \kappa_1 - \kappa_0  }{K(t)} 
\right] 
\label{eqn:pdp:llr:Xstar_only}
\end{align}
which is free of the derivative and the subtraction $(X_T - X_*(t))$. 

However, $L(t)$ in \eqref{eqn:pdp:llr:Xstar_only} is still difficult to realise by chemical reactions because $L(t)$ can be positive and negative numbers. The existing method to implement chemical-based computation system that uses both positive and negative numbers is complex because such systems need more chemical species and reactions \cite{Oishi:2011ig}\cite{Chou:2017bx}. Our proposal to overcome this problem is the same as that in our earlier work \cite{Chou:2018jh}
\cite{Chou:PRE:2022} which is to avoid computing the negative $L(t)$. Our proposal is to compute an approximation $\widehat{L}(t)$ which has the properties:
\begin{itemize}
\item $\widehat{L}(t) \approx L(t)$ when the transmitter sends Symbol 1; and,
\item $\widehat{L}(t) \approx 0$ when the transmitter sends Symbol 0.
\end{itemize}
Since $L(t)$ is positive with a high probability if the transmitter sends a Symbol 1, this means that $\widehat{L}(t)$ is highly likely to be positive when Symbol 1 is sent; and, by the above construction, $\widehat{L}(t)$ is 0 when Symbol 0 is sent. Therefore, we can use $\widehat{L}(t)$ together with a positive threshold to differentiate whether the transmitter sends a 0 or 1. Our proposal to obtain $\widehat{L}(t)$ is to apply the $[\;]_+$ operator to both sides of \eqref{eqn:pdp:llr:Xstar_only} to obtain: 

\begin{align}
\frac{d\widehat{L}(t)}{dt} = & 
g_- X_*(t) 
\underbrace{
\left[
\log \left( \frac{\kappa_1}{\kappa_0} \right) - H_{M0} \Omega \frac{ \kappa_1 - \kappa_0  }{K(t)}  
\right]_+ }_{ \mbox{\scriptsize Threshold-hyperbolic function } \phi(K(t)) } 
\label{eqn:pdp:llr_approx}
\end{align}
where $\widehat{L}(0) = 0$. It can readily be seen that $\widehat{L}(t) \geq 0 \; \forall t \geq 0$. The threshold-hyperbolic (TH) function in \eqref{eqn:pdp:llr_approx} has the property that if $K(t) \leq H_{M0} \Omega  (\kappa_1 - \kappa_0 ) / \log\left( \frac{\kappa_1}{\kappa_0}  \right)$, then $\widehat{L}(t) = 0$. Intuitively, if the transmitter sends a Symbol 0, then the number of signalling molecules $K(t)$ in the receiver voxel is low and consequently the TH function is zero and $\widehat{L}(t) = 0$. In the contrary, if the transmitter sends a Symbol 1, $K(t)$ is large and we have $\widehat{L}(t) > 0$. We will discuss in Sec.~\ref{sec:pdp:circuit} how \eqref{eqn:pdp:llr_approx} can be realised by an enzymatic circuit. In particular, we will make use of the fact that an enzymatic cycle with appropriate rate constants can be used to realise a threshold-hyperbolic function. 


%

\subsection{An enzymatic circuit that approximately calculates log-likelihood ratio} 
\label{sec:pdp:circuit}
In this section, we will explain how an enzymatic circuit can be used to approximately compute $\widehat{L}(t)$ in \eqref{eqn:pdp:llr_approx}. This circuit is located in the receiver voxel. We can see from \eqref{eqn:pdp:llr_approx} that the computation of $\widehat{L}(t)$ requires $X_*(t)$, which comes from the receiver front-end, and $K(t)$ which is the number of signalling molecules in the receiver voxel at time $t$. In particular, the quantity $K(t)$ is used to compute the TH function in \eqref{eqn:pdp:llr_approx}. We will divide our explanation into 3 parts: (i) Using an enzymatic cycle to realise the TH function in \eqref{eqn:pdp:llr_approx}; (ii) Computing the RHS of \eqref{eqn:pdp:llr_approx}; and (iii) Computing the integration in \eqref{eqn:pdp:llr_approx}. 

\subsubsection{Realising the threshold-hyperbolic function}
\label{sec:Lhat:TH}
Our aim is to show how we can use an enzymatic cycle to realise the TH function in \eqref{eqn:pdp:llr_approx} which is a function of $K(t)$. We propose to use the following enzymatic cycle, which will be referred to as the TH-cycle, to realise the TH function: 
\begin{subequations}
\label{cr:p:Z1:all}
\begin{align}
\cee{
Y + K &  <=>[a_1][d_1] YK ->[k_1] Y_* + K  \label{cr:p:Z1:1}  \\
Y_* + P & <=>[a_2][d_2] Y_*P  ->[k_2]  Y + P \label{cr:p:Z1:2}  
} 
\end{align}
\end{subequations}
where \ce{K} is again to be interpreted as the signalling molecules in the receiver voxel whose count at time $t$ is given by $K(t)$. The reactions in \eqref{cr:p:Z1:1} are similar to those in \eqref{cr:p:z0:1} and \eqref{cr:p:z0:2} except that the TH-cycle has a different substrate \ce{Y}. The reactions in \eqref{cr:p:Z1:1} can be used to switch \ce{Y} to its active state \ce{Y_*}, whereas those in \eqref{cr:p:Z1:2} can be used to revert \ce{Y_*} to \ce{Y}. If \eqref{cr:p:Z1:2} is to be interpreted as a dephosphorylation reaction, then \ce{P} is a phosphatase, \ce{Y_*} is the phosphorylated (or active) form, and  \ce{Y_*P} is a complex. The quantities $a_1$ etc. are reaction rate constants. Let $H_{M1} = \frac{d_1 + g_1}{a_1}$ and $H_{M2} = \frac{d_2 + g_2}{a_2}$ be the Michaelis-Menten constants of the TH-cycle. 

We assume that the species \ce{Y}, \ce{YK}, \ce{Y_*}, \ce{P} and \ce{Y_* P} are only found in the receiver voxel and they stay within the voxel. Since the signalling molecule \ce{K} can bind to \ce{X} in the front-end (which is also located in the receiver voxel) as well as to \ce{Y} in the TH-cycle, so the species \ce{X} and \ce{Y} \textsl{compete} for \ce{K} in the receiver voxel. Here, we will analyse the TH-cycle assuming that the front-end \eqref{cr:p:z0:all} is not present. We will consider the interaction of the front-end and the TH-cycle in Sec.~\ref{sec:Lhat:X_times_Y}. 

The \ce{P} molecule can exist in its free form \ce{P} or in the complex \ce{Y_*P}, so the sum of the counts of free form \ce{P} and complex \ce{Y_*P} is a constant, and we will denote it by $P_T$. Similarly, the total count of \ce{Y} in its various forms is denoted by $Y_T$. 

Since the TH-cycle has multiple non-linearities, we carry out a simplified analysis which assumes that the total number of \ce{K} molecules in the receiver voxel is a constant value denoted by $K_T$. We will justify this simplification later on using time-scale separation. 

In 
\ifarxiv
Appendix \ref{app:hyperbolic}, 
\fi
we analyse how the steady-state count of \ce{Y_*}, which is the ``output" of TH-cycle \eqref{cr:p:Z1:all}, depends on $K_T$. Under the assumption that $H_{M1} \gg \frac{K_T}{\Omega}$, $H_{M2} \ll \frac{P_T}{\Omega}$  and sufficiently large $\frac{k_1 K_T}{k_2 P_T}$, we show in 
\ifarxiv
Appendix \ref{app:hyperbolic} 
\fi
using the results in \cite{Straube.2017sal} that the steady state number of \ce{Y_*} molecules can be approximated by the following TH function:
\begin{eqnarray}
Y_* & \approx & 
\left\{
\begin{array}{ll}
 0  &   \mbox{ for } K_T < \frac{h_1 \Omega}{h_0} \\
h_0 - \frac{h_1 \Omega}{K_T}   &  \mbox{ for } K_T > \frac{h_1 \Omega}{h_0} 
\end{array}
\right.
\label{eqn:pdp:hyperbolic} 
\end{eqnarray}
where $h_0 = Y_T - (1 + \frac{k_2}{k_1}) P_T$ and $h_1 = \frac{k_2}{k_1} P_T H_{M1}$. This shows that we can use the steady state number of \ce{Y_*} molecules to realise a TH function of $K_T$. 

We note in particular that a requirement for the TH-cycle \eqref{cr:p:Z1:all} to behave as a TH-function is $H_{M1} \gg \frac{K_T}{\Omega}$, which means $H_{M1}$ has to be chosen so that it is much higher than the concentration of the \ce{K} molecules in the receiver voxel. \rev{We can show that the condition $H_{M1} \gg \frac{K_T}{\Omega}$ implies that few \ce{K} molecules will be sequestered by the TH cycle.} 

We explain in 
\ifarxiv
Appendix \ref{app:TH:para}
\fi
 how we can find reaction rate constants ($a_1$, $d_1$, etc) for the TH-cycle such that the number of \ce{Y_*} molecules is proportional to the TH function in \eqref{eqn:pdp:llr_approx}. Note that the choice of the rate constants has to make $H_{M1} \gg \frac{K_T}{\Omega}$, $H_{M2} \ll \frac{P_T}{\Omega}$, which are required for the cycle in \eqref{cr:p:Z1:all} to behave as a TH-function. In particular, given that $H_{M1} = \frac{d_1 + g_1}{a_1}$ has to be large, this can be achieved by having a small $a_1$ which implies that the reaction \ce{Y + K -> YK} is slow. \rev{According to \cite{Cao:2005gj}, this slow binding rate means this reaction is then driven by the average number of \ce{K} molecules.
}  
 At the same time, the calculation of the TH-function by \eqref{eqn:pdp:hyperbolic} is based on the steady state. Overall, this means the calculation of the TH-function by the TH-cycle \eqref{cr:p:Z1:all} is driven by the steady state mean number of $K(t)$. This justifies why we could use a constant $K_T$ for an approximate analysis in 
\ifarxiv
Appendix \ref{app:hyperbolic}. 
\fi

\subsubsection{Computing the RHS of \eqref{eqn:pdp:llr_approx}}
\label{sec:Lhat:X_times_Y}
In this section, we will explain how we can compute the RHS of \eqref{eqn:pdp:llr_approx}. 

We first address the issue that both the receiver front-end (with substrate \ce{X}) and the TH-cycle (with substrate \ce{Y}) compete for signalling molecule \ce{K} in the receiver voxel to produce the active form \ce{X_*} and \ce{Y_*} in their respective cycle. So far, we have analysed the front-end and the TH-cycle separately in Sec.~\ref{sec:pdp:approx:front_end:para} and \ref{sec:Lhat:TH}, without considering the competition for \ce{K} between these cycles. As an illustration of this competition, let us assume that some \ce{K} molecules have been bound to form \ce{XK} in the front-end, then these \ce{K} molecules are no longer available for binding with \ce{Y} in the TH-cycle and this means the TH-cycle 
``sees" a reduced number of \ce{K} molecules. 
\rev{
Fortunately, both the front-end and the TH-cycle have high Michaelis-Menten constants $H_{M0}, H_{M1} \gg \frac{K_{1}^{\rm ss}}{\Omega}$, hence very few \ce{K} molecules are sequestered in \ce{XK} and \ce{YK}. 
}
Therefore, we can justify the derivations earlier where we analyse the front-end and the TH-cycle separately. 

We now explain how the RHS of \eqref{eqn:pdp:llr_approx} can be computed. 
From the design of the TH-cycle, we can make the TH-function in \eqref{eqn:pdp:llr_approx} to be proportional to $Y_*(t)$, so the RHS of \eqref{eqn:pdp:llr_approx} is proportional to the product $X_*(t) Y_*(t)$. We propose to realise this product by using a molecule that has two binding sites to which the  signalling molecule \ce{K} can bind. We use the symbol \ce{X-Y} to denote the structure of this molecule where the \ce{X} part of the molecule behaves like the \ce{X} in the front-end \eqref{cr:p:z0:all} and similarly the \ce{Y} part behaves as the \ce{Y} in the TH-cycle \eqref{cr:p:Z1:all}. The molecule \ce{X-Y} can have numerous states, e.g., \ce{X-Y} (where both \ce{X} and \ce{Y} are inactive), \ce{X_*-Y},\ce{X-Y_*}, \ce{X_*-Y_*} (where both \ce{X} and \ce{Y} are active) etc. With the assumption that the probabilities of \ce{X} and \ce{Y} being activated by \ce{K} are independent, we show in 
\ifarxiv
Appendix \ref{app:Xp_times_Yp}
\fi
 that the product $X_*(t) Y_*(t)$ is proportional to the number of molecules in the \ce{X_*-Y_*} state. Therefore, a scaled version of $\widehat{L}(t)$ in \eqref{eqn:pdp:llr_approx} can be obtained from integrating the number of molecules in \ce{X_*-Y_*} state over time. 
As a remark, we want to point out that there are many examples of protein molecules that have multiple binding sites. For example, if the binding results in phosphorylation, then these types of molecules are studied under the topic of multi-site phosphorylation \cite{Salazar.2009}. We also note that independent binding assumption is also used in a lot of analysis in multi-site phosphorylation \cite{Salazar.2007}; this justifies the use of independent binding assumption earlier. 

We have so far not discussed the time-scale for the reactions in the enzymatic cycles for \ce{X} and \ce{Y}. We see from our earlier analysis that the TH-function calculation in \eqref{eqn:pdp:hyperbolic} requires the TH-cycle to reach the steady state. This means that only the steady state portion of the TH-cycle signal is useful; this also holds for the receiver front-end. Therefore, in order to achieve accurate detection, we will need the integration time for computing $\widehat{L}(t)$ to be long compared to the time-scale for the two enzymatic cycles for \ce{X} and \ce{Y} to reach the steady state. In other words, both enzymatic cycles should have a short transient compared with the symbol duration. 

\subsubsection{Computing the integration in \eqref{eqn:pdp:llr_approx}}

In this paper, we will use an enzymatic cycle to realise the integration in \eqref{eqn:pdp:llr_approx}. We assume that \ce{X_*-Y_*} reacts with a substrate \ce{J} in the following enzymatic cycle: 
\begin{subequations}
\label{cr:p:int:all}
\begin{align}
\cee{
J + X_*-Y_* &  <=>[a_3][d_3] JX_*-Y_* ->[k_3] J_* + X_*-Y_*  \label{cr:p:int:1}  \\
J_* + \tilde{P} & <=>[a_4][d_4] J_*\tilde{P}  ->[k_4]  J + \tilde{P} \label{cr:p:inr:2}  
} 
\end{align}
\end{subequations}
Our aim is to ensure that the number of \ce{J_*} molecules is proportional to the integral $\int_0^t \; \{ X_*-Y_* \}(\tau) \; d\tau$ and therefore proportional to $\hat{L}(t)$. We achieve this by using large Michaeis-Menten constants $K_{M3} = \frac{d_3 + k_3}{a_3}$ and $K_{M4} = \frac{d_4 + k_4}{a_4}$in both directions of the cycle and to ensure that the cycle operates far away from saturation. The use of a large Michaelis-Menten constant for the forward path \eqref{cr:p:int:1} of the cycle also ensures that few \ce{X_*-Y_*} are sequestered. Since the design of the integrator is not a key purpose of this paper, we will not delve into further details. 

\subsubsection{Enzymatic receiver circuit} 
We have now completed the description of the three enzymatic cycles ---  front-end \eqref{cr:p:z0:all}, TH-cycle \eqref{cr:p:Z1:all} and integrator \eqref{cr:p:int:all} --- which work together to compute the approximate log-probability ratio $\widehat{L}(t)$. These three cycles together form an enzymatic cycle-based receiver. We will use the term \textsl{enzymatic receiver} to refer to the receiver formed by these 3 cycles. 

\section{Numerical evaluations}
\label{sec:eval} 
The aim of this section is to use numerical experiments to understand the properties of the proposed enzymatic receiver. We first describe the experimental settings. 

\subsection{Experimental settings}  
\label{sec:eval:set_up} 
\begin{figure}[t]
        \centering
        \includegraphics[scale = \picscale]{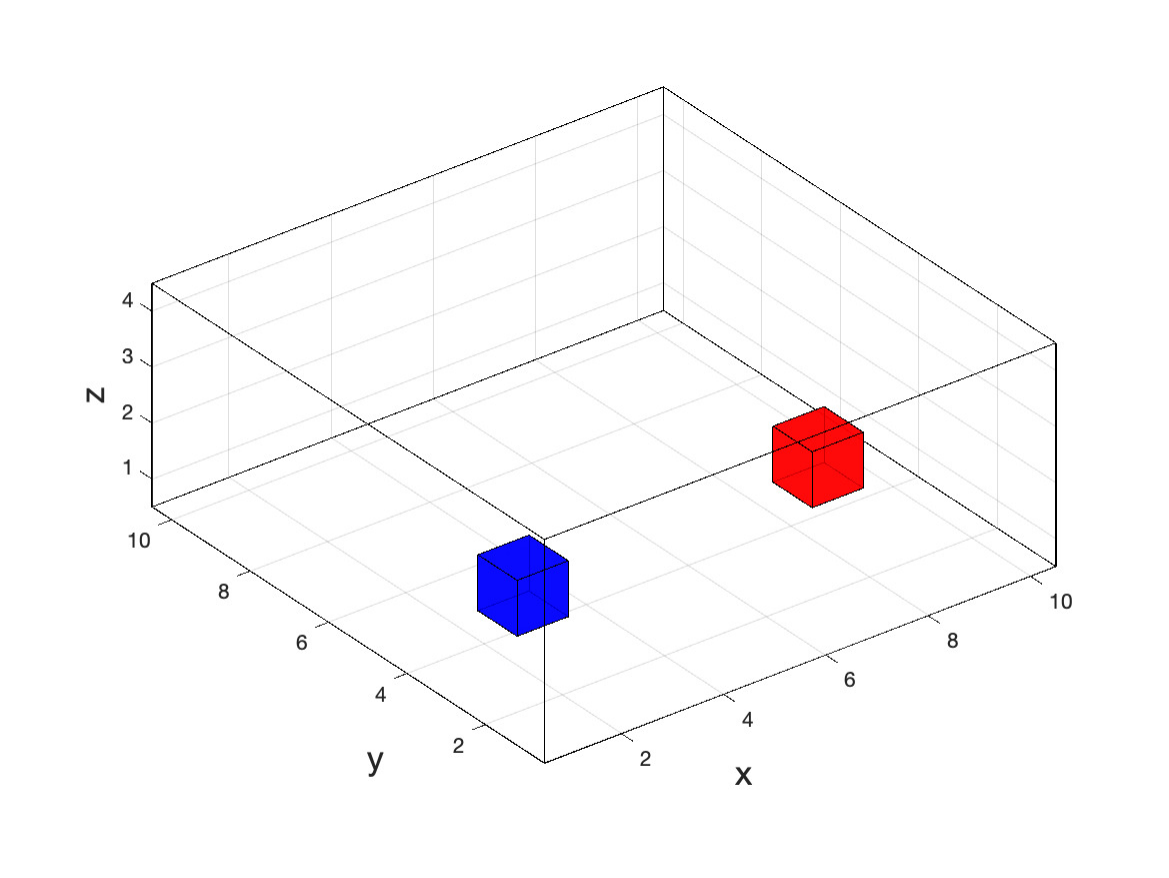} 
\caption{\rev{Location of the transmitter (blue) and receiver (red) voxels.}}      
\label{fig:voxel_locations}
\end{figure}

\rev{
We consider a medium of 10$\mu$m $\times$ 10$\mu$m $\times$ 4 $\mu$m. We assume a voxel size of $W^3$ $\mu$m$^{3}$ where $W = 1$, creating an array of $10 \times 10 \times 4$ voxels.  The voxel coordinates for the transmitter and receiver are, respectively, (2,3,2) and (7,2,3). Fig.~\ref{fig:voxel_locations} illustrates the location of the voxels.}

\rev{We assume the diffusion coefficient $D$ of the vesicles in the medium is \rev{0.1} $\mu$m$^2$s$^{-1}$. This is within the range of the diffusion coefficient for extra-cellular vesicles in \cite{Kastelowitz:2014}.} We assume an absorbing boundary for the medium and the \rev{vesicles} escape from a boundary voxel surface at a rate of $\frac{D}{50W^2}$. 

\rev{
We assume that the transmitter uses the following reactions to produce the signalling molecules and vesicles: 
\begin{subequations}
\label{cr:trans:all}
\begin{align}
\cee{mRNA &->[r_{\rm K}] mRNA + K}  \label{cr:trans:K:prod}
\\
\cee{K &->[r_{\rm d}] \phi}  \label{cr:trans:K:degrade}
\\
\cee{K &->[r_{\rm v}] \hat{K} } \label{cr:trans:V:prod}
\end{align}
\end{subequations}
where we assume that the signalling molecules \ce{K} is produced in the transmitter via a translation reaction at a rate of $r_{\rm K}$. Reaction \eqref{cr:trans:K:degrade} says that \ce{K} may degrade. Finally, \eqref{cr:trans:V:prod} captures the encapsulation of \ce{K} in a vesicle and its emission. We choose $r_{\rm K}$ to be 0.1932 proteins/mRNA/s which is below the translation rate of 1000 proteins/mRNA/hour in \cite{10.1038/nature11848}. The parameter $r_{\rm v}$ is chosen to be 0.1 vesicles per second which is below the production rate of 1 vesicle/cell/s in \cite{Woehler:EV:2024}. We assume that the two symbols are produced by using different number of mRNA molecules. In the absence of the receiver reactions, the steady state counts \ce{K} molecules in the receiver voxel are respectively 12 and 40 molecules when Symbols 0 and 1 are sent. Lastly, we remark that the above reactions take place in the transmitter voxel, and that both mRNA and \ce{K} cannot leave the transmitter voxel, only $\hat{\rm K}$ can. 
}


\rev{
The receiver front-end has 4 parameters: $a_0$, $d_0$, $g_0$, $g_-$ where $a_0$ is a bimolecular reaction rate constant while the other three are unimolecular. We use the paper \cite{Takahashi:2010ko} to guide us to choose the unimolecular rates. The unimolecular rates in \cite{Takahashi:2010ko} are in the range 1.35-15. We have chosen $d_0 = g_0 = 8$ which are within this range. We will use different values of $H_{M0}$ and $\gamma$ to study their impact on the BER. The different values of $H_{M0}$ that we will use is within the range of Michaelis-Menten constants of 100--10$^8$nM stated in \cite{libretextsMichaelisMentenKinetics}. Given the choices of $H_{M0}$ and $\gamma$, we will calculate $a_0$ using $\frac{d_0 + g_0}{H_{M0}}$ and $g_- = \frac{g_0}{\gamma}$. The number of \ce{X-Y} molecules is chosen to be 60. We use our algorithm in 
\ifarxiv
Appendix \ref{app:TH:para} (which is discussed in Sec.~\ref{sec:pdp:circuit})
\fi
 to calculate the parameters of the TH-cycle. We put the value of the calculated TH-cycle parameters for $H_{M0} = 1000$nM and $\gamma = 28.6$, as well as those of the integrator, in Appendix \ref{app:cycle:para}. We also ensure that all the bimolecular reaction rate constants are within the diffusion-limited binding rate which we have shown in 
\ifarxiv
Appendix \ref{app:diff_limited}
\fi 
to be about 7.5 nM$^{-1}$s$^{-1}$.  
}

\begin{figure}[t]
        \centering
        \includegraphics[scale = \picscale]{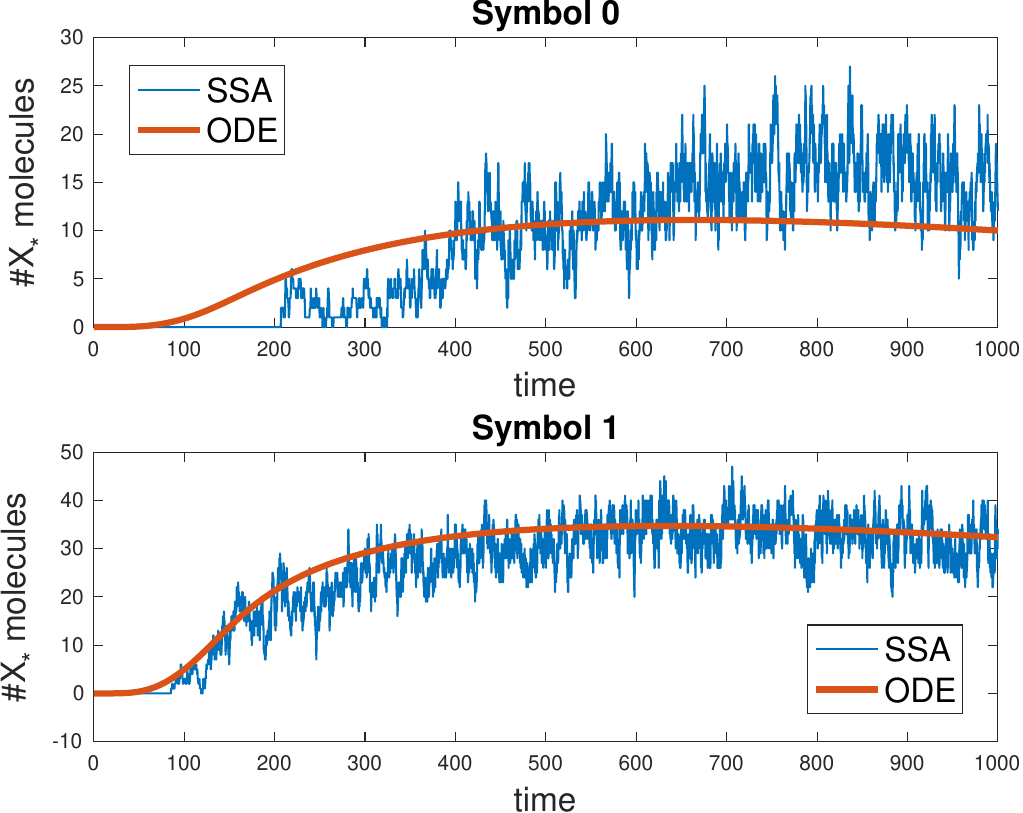} 
\caption{\rev{Number of \ce{X_*} molecules for Symbol 0 and 1.}   }      
\label{fig:eval:Xp_count}
\end{figure}

All the simulations are carried out by using the Stochastic Simulation Algorithm (SSA) \cite{Gillespie:1977ww}. As an example of the time signal, Fig.~\ref{fig:eval:Xp_count} plots the number of \ce{X_*} molecules in the receiver when Symbols 0 and Symbol 1 are sent. 

%

\subsection{Filtering approximation} 
\label{sec:filtering}
The aim of this section is to verify the accuracy of the closed-form formula \eqref{eqn:Js:approx} that approximately computes the posteriori mean ${\rm E}[\{XK\}(t) | s, {\cal X}_*(t)]$ of the number of \ce{XK} molecules. The exact numerical computation of the posteriori mean is demanding because we need to compute the time evolution of the probabilities of the all the states that are compatible with the given observations ${\cal X}_*(t)$. 
\rev{Specifically, the state that we need to include in this filtering problem composes of the count of all species in the transmitter, the counts of vesicles in each voxel and the counts of species in the receiver front-end that cannot be deduced from the observations and conservation. The number of states is enormous for the medium specified in Sec.~\ref{sec:eval:set_up}. If we only count the medium and we assume each voxel can only have 10 vesicles, then we need $11^{398}$ different states just for the medium. 
} 
 We therefore need to use a smaller medium for this verification. We have chosen to use a medium which consists of one voxel, which means the transmitter and receiver are co-located at the same voxel. We will need to calculate the evolution of approximately 2500 state probabilities which means solving 2500 ODEs of the same number of variables. 

\begin{figure}[t]
        \centering
        \includegraphics[scale = \picscale]{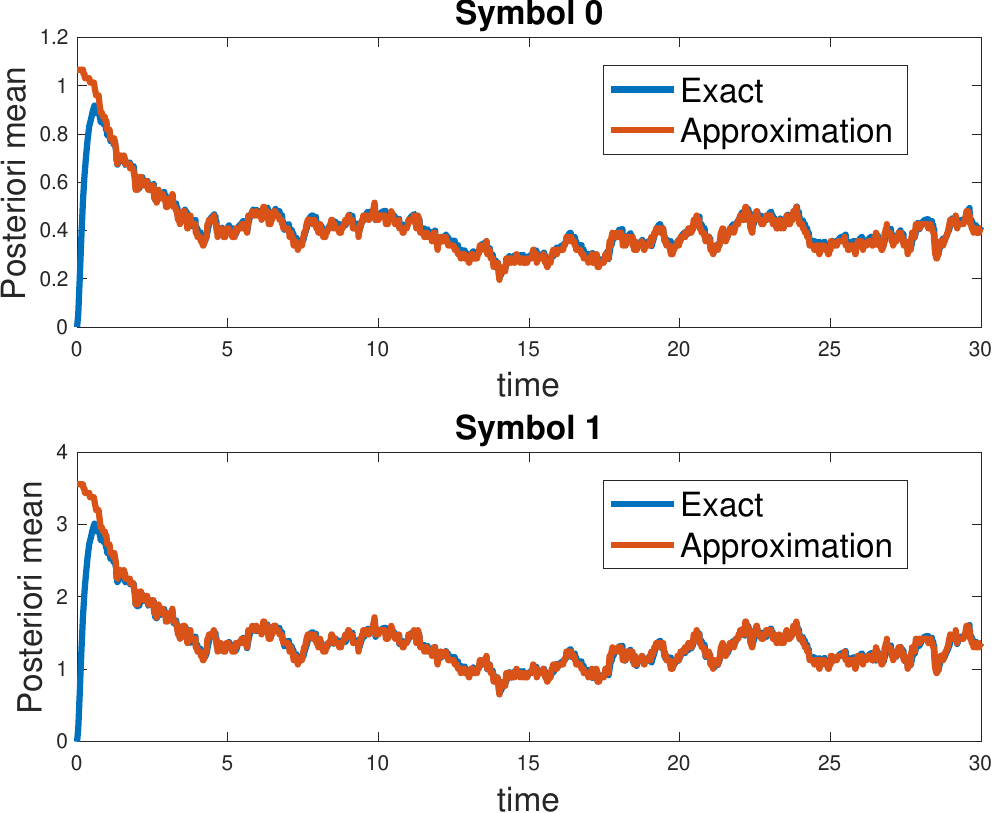} 
\caption{\rev{The exact posteriori mean $J_s(t) =  {\rm E}[\{XK\}(t) | s, {\cal X}_*(t)]$ and its approximation is \eqref{eqn:Js:approx}. } }      
\label{fig:eval:filtering}
\end{figure}

We use the 1-voxel medium, \rev{the transmitter} and the receiver front-end parameters together with SSA algorithm to simulate the system. We use the system model and the observations ${\cal X}_*(t)$ to compute the exact posteriori mean and its closed-form approximation. Fig.~\ref{fig:eval:filtering} shows the results for both Symbols 0 and 1. It can be seen that the approximation is accurate except for the initial transient. This is because the approximation assumes that the system is in steady state so it cannot deal with the transient. 

\subsection{Log-probability approximation - small medium} 
\label{sec:eval:llr:small_medium}
The aim of this section is to demonstrate that the approximate log-probability ratio computed by \eqref{eqn:ode:L:kappa} and \eqref{eqn:pdp:llr_approx} are accurate. We compare these approximations against the exact log-probability ratio $L(t)$ in \eqref{eqn:ode:L}. Since the numerical calculation of the exact $L(t)$ requires the solution of a Bayesian filtering problem, so the comparison here is based on the small medium that we use in Sec.~\ref{sec:filtering}. 

\begin{figure}[t]
        \centering
        \includegraphics[scale = \picscale]{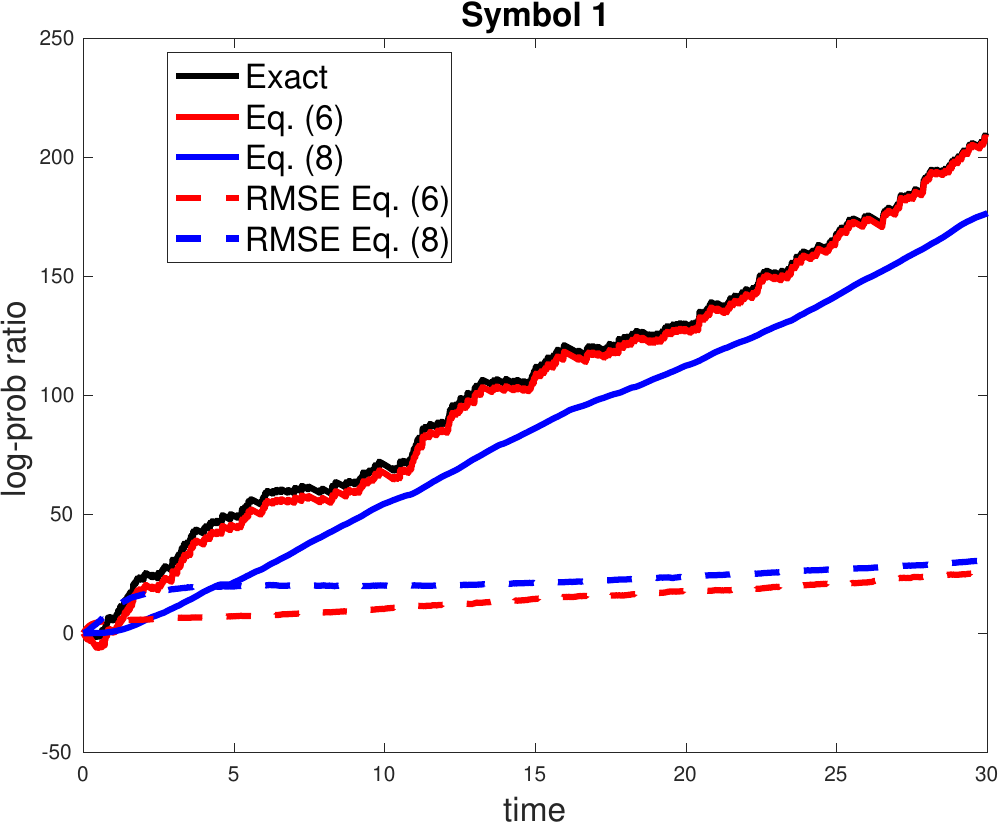} 
\caption{\rev{One realisation of the exact log-probability ratio and its approximation \eqref{eqn:ode:L:kappa} and \eqref{eqn:pdp:llr_approx} (solid line). RMSE between exact and the approximation (dashed lines). Symbol 1.}}      
\label{fig:eval:llr:sym1:small_medium}
\end{figure}

Fig.~\ref{fig:eval:llr:sym1:small_medium} compares, for Symbol 1, one realisation of the exact log-probability ratio $L(t)$
 (in black solid line) and its approximation \eqref{eqn:ode:L:kappa} (in red solid line) and \eqref{eqn:pdp:llr_approx} (in blue solid line). The approximation \eqref{eqn:ode:L:kappa} is very accurate as the red line sits almost on top of the black line. We further compute the root-mean-square error (RMSE) of the two approximations over 200 realisations. The results are plotted as dash lines in Fig.~\ref{fig:eval:llr:sym1:small_medium}. We can see that both approximations have small RMSEs. 

\begin{figure}[t]
        \centering
        \includegraphics[scale = \picscale]{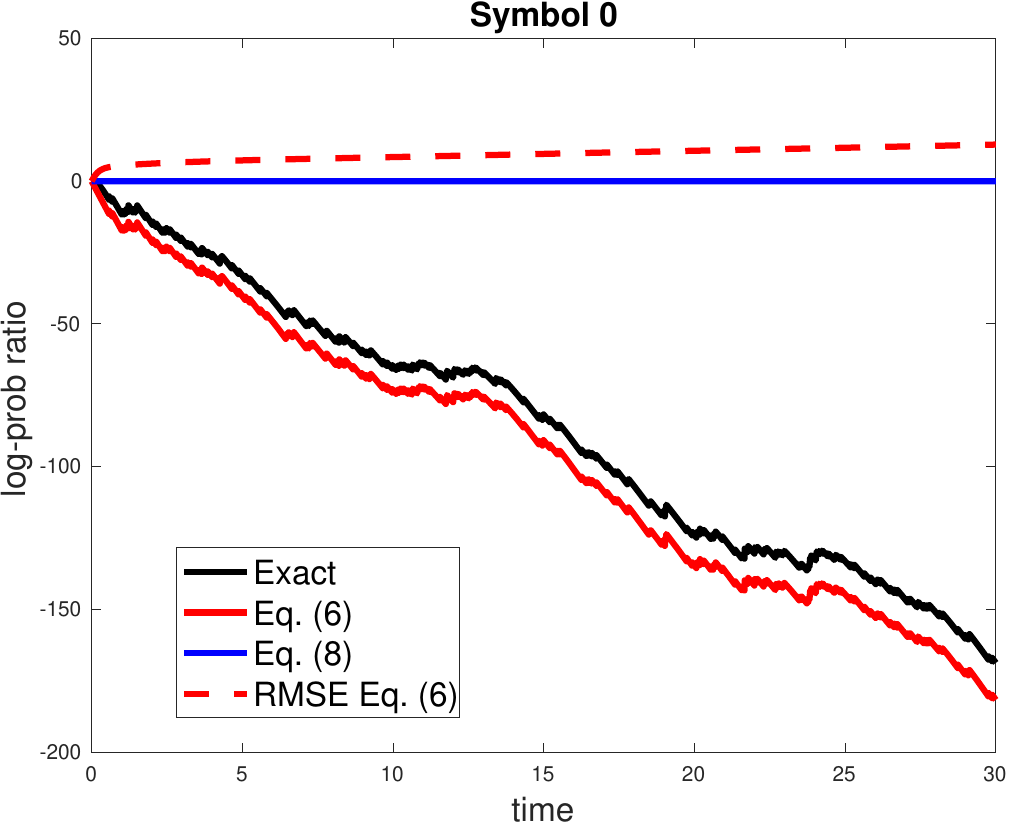} 
\caption{\rev{One realisation of the exact log-probability ratio and its approximation \eqref{eqn:ode:L:kappa} and \eqref{eqn:pdp:llr_approx} (solid line). RMSE between exact and the approximation (dashed lines). Symbol 0.}}      
\label{fig:eval:llr:sym0:small_medium}
\end{figure}

Fig.~\ref{fig:eval:llr:sym0:small_medium} shows similar comparisons but this time for Symbol 0. The approximation \eqref{eqn:ode:L:kappa} (in red solid line) is again very accurate. For Symbol 0, the approximation \eqref{eqn:pdp:llr_approx} is 0 as shown by the blue solid line in the figure. 

Since we have shown that \eqref{eqn:ode:L:kappa} is an accurate approximation of the exact $L(t)$ and we are unable to compute the exact $L(t)$ for large medium, we will use \eqref{eqn:ode:L:kappa} as the correct $L(t)$ for the larger medium defined in Sec.~\ref{sec:eval:set_up}. From this point onwards, the results  will be based on the larger medium. 

\subsection{Log-probability ratio computed by an enzymatic receiver} 
\label{sec:eval:circuit:approx}
\begin{figure}[t]
        \centering
        \includegraphics[scale = \picscale]{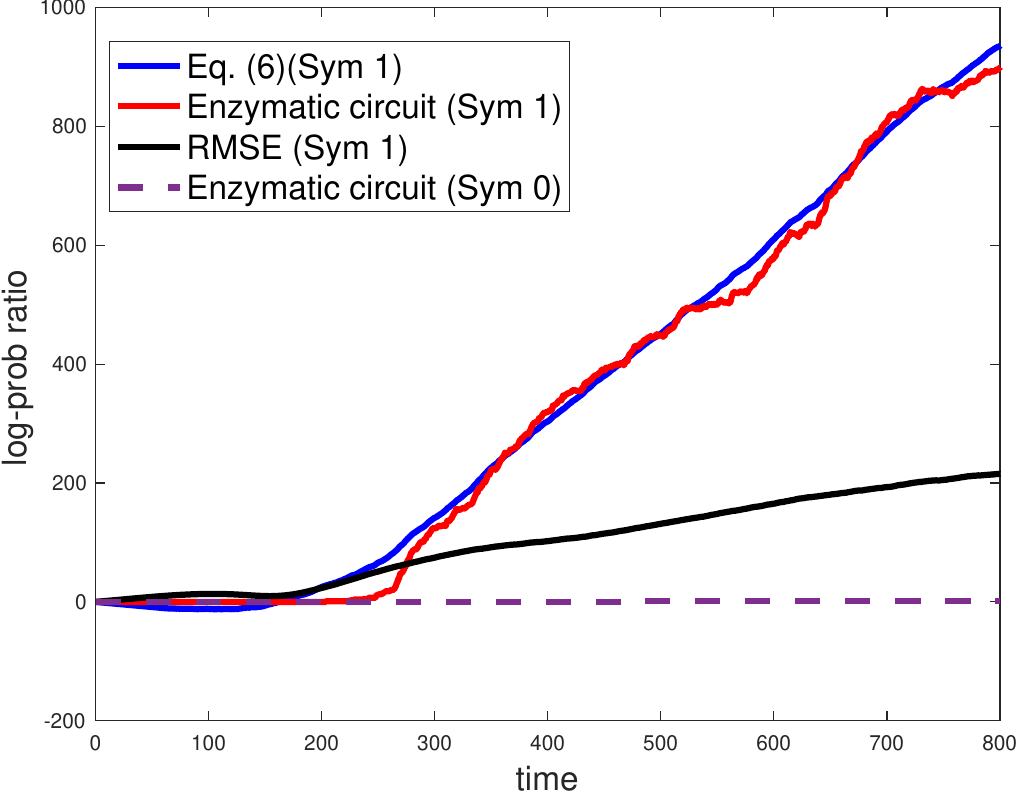} 
\caption{\rev{Comparing the output of the enzymatic circuit against that of log-probability ratio \eqref{eqn:ode:L:kappa}.}}      
\label{fig:eval:circuit:output}
\end{figure}

The aim of this section is to show that the output of the enzymatic receiver is approximately equal to the log-probability ratio when Symbol 1 is sent and is almost 0 when Symbol 0 is sent.  Fig.~\ref{fig:eval:circuit:output} compares, for Symbol 1, a realisation of  the enzymatic circuit output (red line) and that of approximate log-probability ratio \eqref{eqn:ode:L:kappa} (blue line). It can be seen that, after a period of transient, they become fairly close to each other. The black line in Fig.~\ref{fig:eval:circuit:output} shows the RMSE, over 200 realisations, between the enzymatic circuit output and \eqref{eqn:ode:L:kappa}; it can be seen that the RMSE is low. Lastly, when Symbol 0 is sent, the output of the enzymatic circuit is almost zero as shown by the magenta dashed lines in the Fig.~\ref{fig:eval:circuit:output}. 

\subsection{Bit-error rate} 
\label{sec:eval:circuit:ber}
\rev{This section studies the BER of the enzymatic receiver circuit for different values of $H_{M0}$ and $\gamma$ for the receiver front-end.}

\begin{figure}[t]
        \centering
        \includegraphics[scale = \picscale]{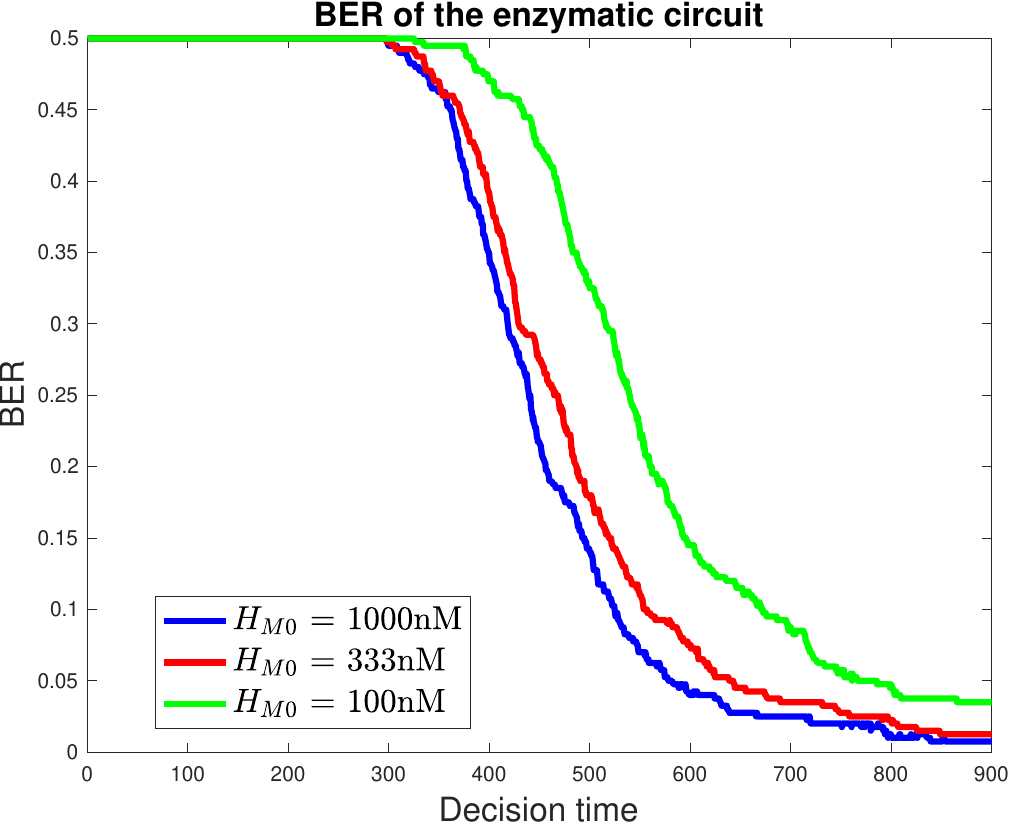} 
\caption{\rev{Comparing the BER for 3 different values of $H_{M0}$: 1000, 333 and 100nM.}}      
\label{fig:eval:circuit:ber:v34:Km}
\end{figure}

\begin{figure}[t]
        \centering
        \includegraphics[scale = \picscale]{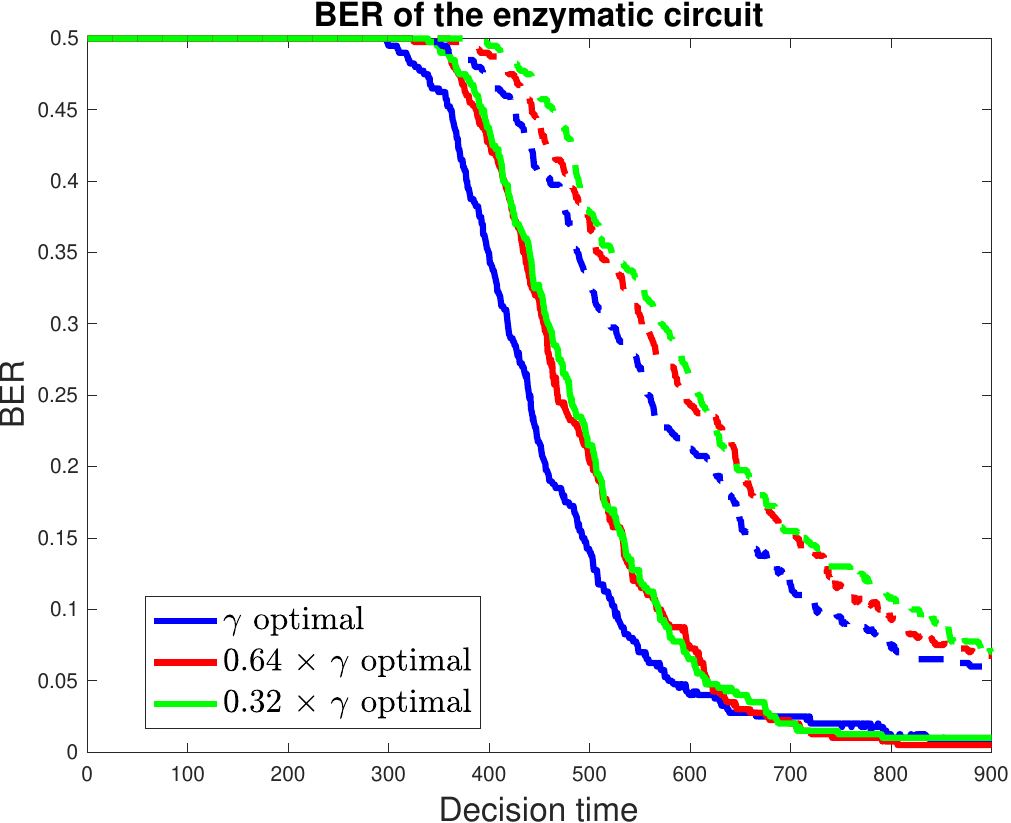} 
\caption{\rev{Comparing the BER for 3 different values of $\gamma$: optimal and 2 sub-optimal values.}}      
\label{fig:eval:circuit:ber:v34:gamma}
\end{figure}

The demodulation decision is based on the output of the enzymatic receiver circuit. If the output is higher than a threshold at the decision time, then the demodulator decides that Symbol 1 has been sent; otherwise it decides for Symbol 0. In this study, we fix the decision threshold to 200 and vary the decision time from 0 to 900. 
\rev{The time to decision is longer because of lower diffusion rate of the vesicles in the medium.
Fig.~\ref{fig:eval:circuit:ber:v34:Km} shows how the BER varies with decision time for $H_{M0} = $ 1000nM, 333nM and 100nM. Note that we have set the steady state concentration $[K]^{\rm ss}_1$ of \ce{K} in the receiver to be 66.4nM so the larger two values of $H_{M0}$ are about 10 and 5 times of $[K]^{\rm ss}_1$. 
We can see that we can achieve a small BER when the decision time is long enough. We can clearly see that a larger $H_{M0}$ value produces a lower BER at a given time. 

Similarly, the solid lines in Fig.~\ref{fig:eval:circuit:ber:v34:gamma} shows the BER for different values of $\gamma$ when $H_{M0}$ is fixed at 1000nM. We find that the optimal $\gamma$ gives a lower BER compared to non-optimal choices. By comparing these two figures, we find that if the value of $H_{M0}$ is already large, the optimal $\gamma$ is only slightly better than non-optimal ones. 

We also study the robustness of the receiver circuit to perturbations in the transmitter circuit used to produce the two transmission symbols. 
In the transmitter reactions in \eqref{cr:trans:all}, we assume Symbols 0 and 1 are produced by using $N_0$ and $N_1$ number of mRNA molecules where $N_1 > N_0$.  We consider a perturbation scenario where $N_0$ (resp. $N1$) is increased by 
100\% (resp. reduced by 17\%). This will make the symbols harder to distinguish. The dashed lines in Fig.~\ref{fig:eval:circuit:ber:v34:gamma} show the BER, for different values of $\gamma$, under this perturbation. We find that the enzymatic circuit can still demodulate but it takes a longer time to reach a decision. 
}



\ifarxiv 
We have also studied a number of variations in terms of medium dimension, transmitter type, symbol duration, different $H_{M0}$ values. 

Figs.~\ref{fig:eval:circuit:ber:v14:Km} and \ref{fig:eval:circuit:ber:v14:gamma} show the impact of $H_{M0}$ and $\gamma$ for an alternative medium dimension. The medium has a smaller dimension of 6 $\mu$m $\times$ 6 $\mu$m $\times$ 3 $\mu$m. We also used different $H_{M0}$ values of 1000, 333 and 100nM. Other than these two adjustments, the other parameters remain the same. We can see that both figures show the same type of parameter impact on performance as before. 

\begin{figure}[t]
        \centering
        \includegraphics[scale = \picscale]{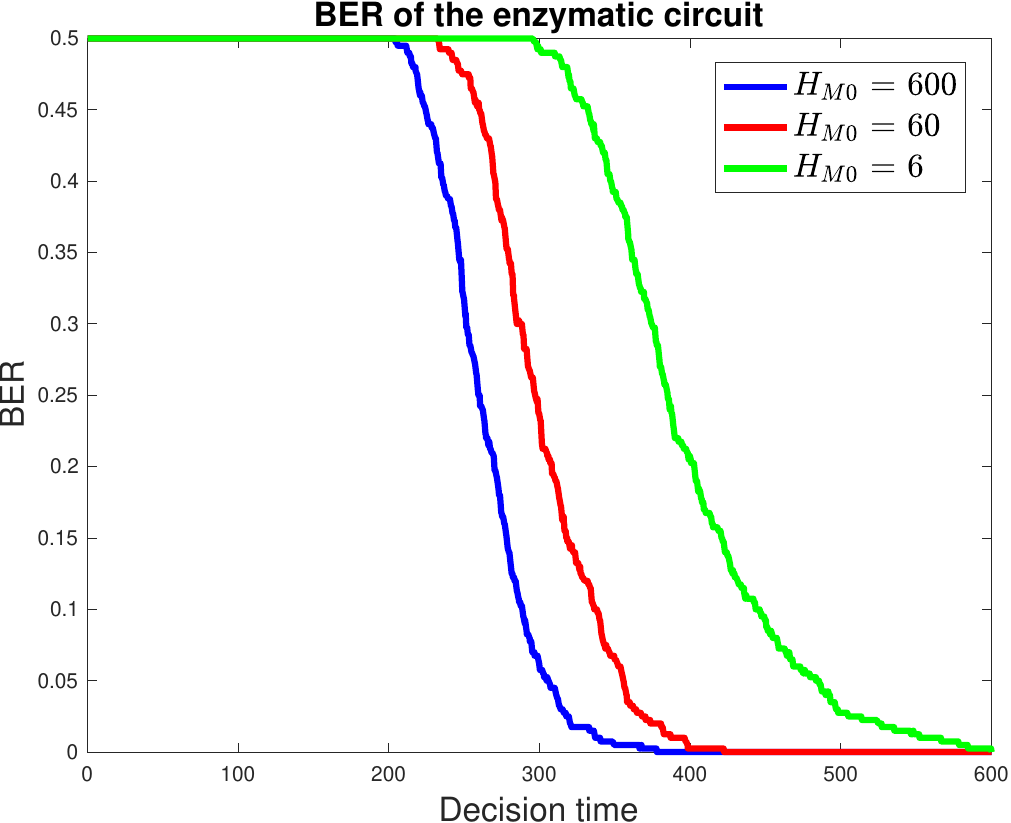} 
\caption{\rev{Comparing the BER for 3 different values of $H_{M0}$: 1000, 333 and 100.} Based on an alternative medium dimension.}      
\label{fig:eval:circuit:ber:v14:Km}
\end{figure}

\begin{figure}[t]
        \centering
        \includegraphics[scale = \picscale]{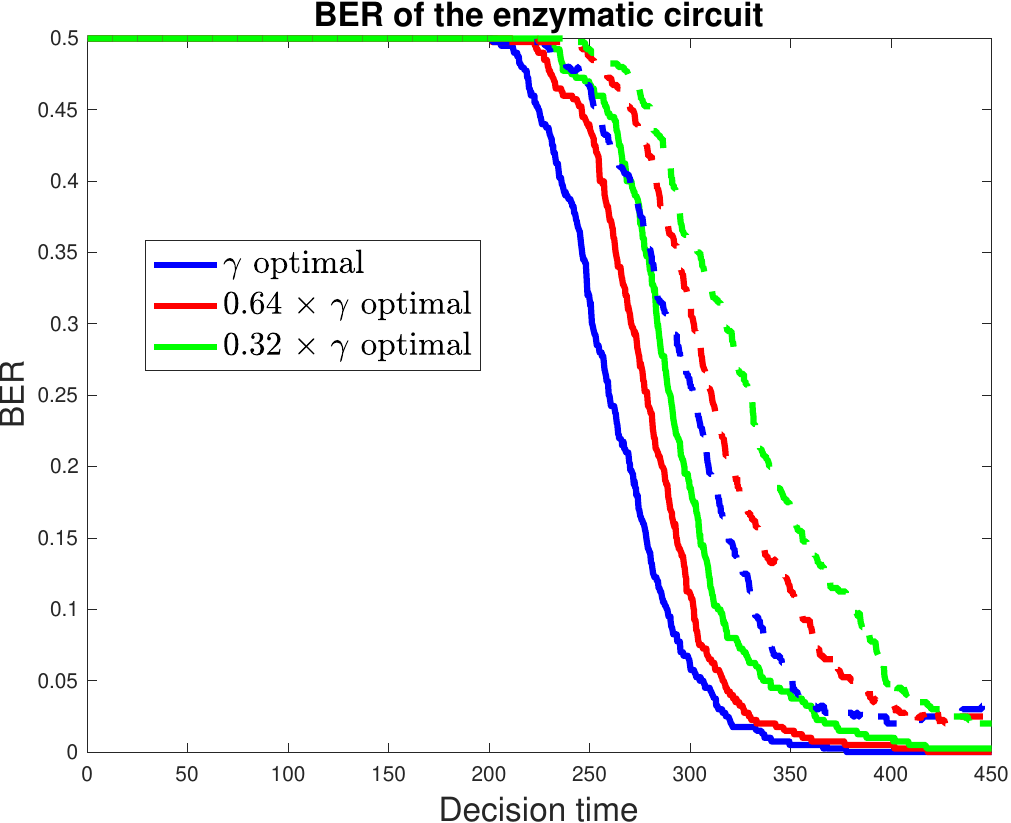} 
\caption{\rev{Comparing the BER for 3 different values of $\gamma$: optimal and 2 sub-optimal values.} Based on an alternative medium dimension.}      
\label{fig:eval:circuit:ber:v14:gamma}
\end{figure}

Continuing on using an alternative medium dimension, we will now use an alternative transmitter model. For this model, we only model the emission of the vesicle into the medium and it assumes that this emission is a Poisson process. This emission rates for Symbols 0 and 1 will again produce 12 and 40 signalling molecules in the receiver in the absence of the receiver circuit. Therefore, the receiver sees the same concentration shift keying rate but the transmitter process is different. We have also reduced the symbol duration to 30. Figs.~\ref{fig:eval:circuit:ber:v12:Km} and \ref{fig:eval:circuit:ber:v12:gamma} show the impact of $H_{M0}$ and $\gamma$ for this alternative transmitter in the alternative medium. 

We have also varied the positions of the transmitter and receiver voxels. For example, earlier we used a receiver voxel at $(7,3,2)$, we have also used $(8,2,3)$, $(3,5,2)$, $(4,5,2)$, $(3,4,3)$. We have also varied the parameters of the transmitter circuit by changing its $r_K$ and $r_v$ parameters, duration of the symbols, and the number of mRNA used to produced the symbols. We found that the results are similar to what we presented earlier. The key results, which we have illustrated earlier, are that higher Michaelis-Menten constant $H_{M0}$ and optimal value of $\gamma$ are useful for reducing the BER.

\begin{figure}[t]
        \centering
        \includegraphics[scale = \picscale]{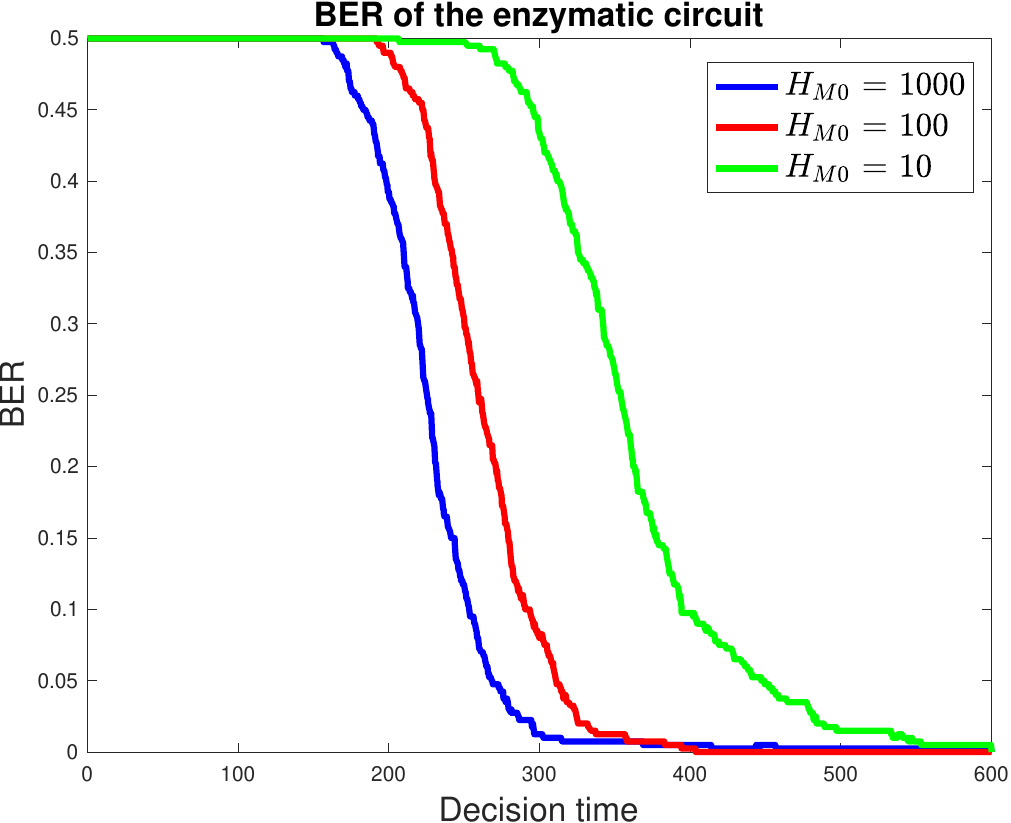} 
\caption{\rev{Comparing the BER for 3 different values of $H_{M0}$: 1000, 100 and 10nM.} Based on an alternative medium dimension and transmitter.}      
\label{fig:eval:circuit:ber:v12:Km}
\end{figure}

\begin{figure}[t]
        \centering
        \includegraphics[scale = \picscale]{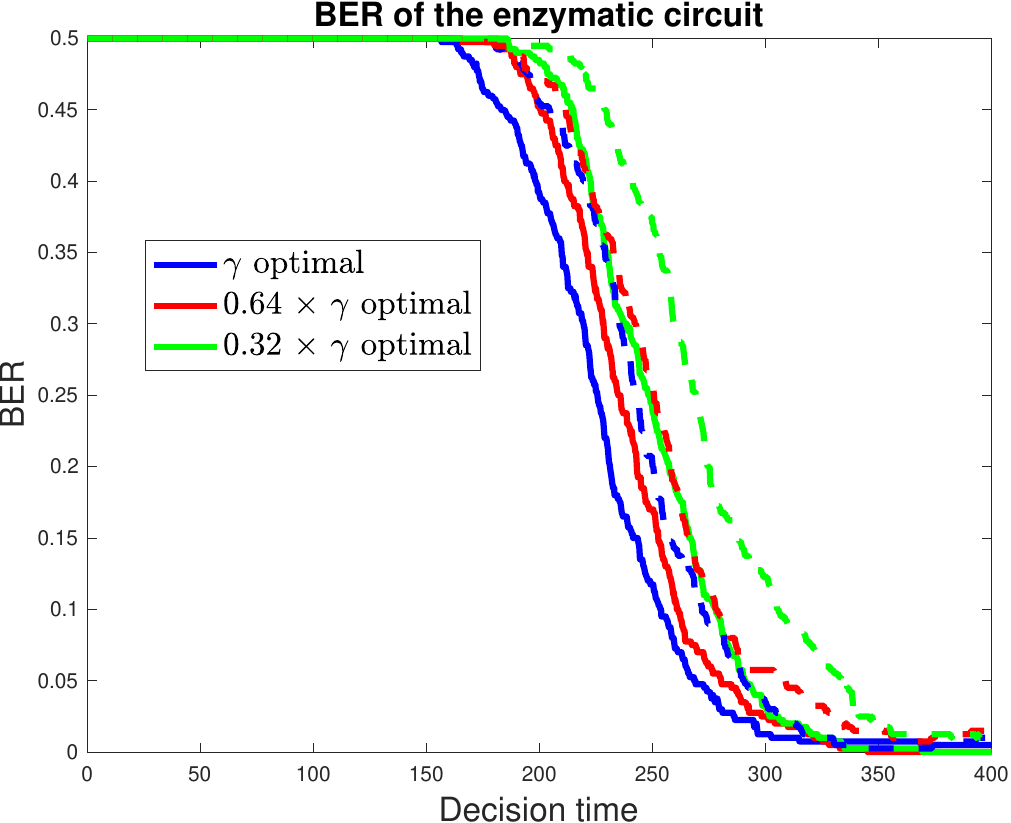} 
\caption{\rev{Comparing the BER for 3 different values of $\gamma$: optimal and 2 sub-optimal values.} Based on an alternative medium dimension and transmitter.}      
\label{fig:eval:circuit:ber:v12:gamma}
\end{figure}

\fi

%

\section{Discussions}
\rev{
We noted in Sec.~\ref{sec:intro} that this paper is based on our earlier work \cite{Awan:2017fm} which assumes the architecture in Fig.~\ref{fig:system_overview}. There, we also pointed out that the work in \cite{Awan:2017fm} is general and can be used to derive the ODE for calculating the log-posteriori probability (similar to \eqref{eqn:ode:L}) in the back-end for for \textsl{any} chemical-reaction based modulation scheme in the transmitter and for \textsl{any} front-end molecular circuits. One may wonder whether it is possible to derive reaction-based demodulator for other reaction-based modulation methods and front-end circuits. Based on the work in this paper, we see that there two open problems. 

First, we need to determine an approximate solution to the filtering problem associated with MAP detection. Although it is possible to write down the Bayesian filtering solution as a set of ODE and discrete jumps (see \cite{Bronstein:2018eh} and Appendix \ref{app:filtering}), finding an approximation to such a set of equations is still an open problem. For concentration shift keying, it is sufficient for us to find a first order approximation but for other modulation schemes, one may need to find a lower order dynamical system as the approximation and this is challenging. 

Second, we need to find a set of chemical reactions that can produce certain dynamical behaviour as the demodulator. Although there is some existing work that can achieve such goals \cite{Lormeau:NatureComm:2021}, such work assumes that the types of reactions have been chosen which means one must ensure that the optimisation space is large enough. The challenge here is to derive an efficient solver that can deal with the specific combinatorial optimisation problem for reaction network design. 

One may also ask how general the front-end and back-end architecture in Fig.~\ref{fig:system_overview} is. The derivation in \cite{Awan:2017fm} assumes that the back-end does not significantly sequester the output molecules from the front-end. This can be done if the front-end and back-end are isolated or the connection has low retro-activity \cite{McBride:PIEEE:2019}. There is some work on designing chemical circuits to enable this isolation \cite{Mishra:2014da} so the division into front-end and back-end may not pose severe problems. 

}

\section{Conclusions} 
\label{sec:con} 

\rev{
This paper presents an enzymatic-circuit based receiver that can approximately  demodulate concentration modulated signals using the MAP criterion. We address a number of problems to realise this goal. We begin by studying how the parameter of an enzymatic circuit can be chosen to enhance its sensitivity to concentration modulated symbols. We then derive a closed-form formula for a filtering problem in the MAP demodulation problem and then derive an approximate ODE for calculating the log-probability ratio. We finally show that this approximate ODE can be realised by an enzymatic circuit. An interesting finding is that, for the chosen  enzymatic front-end, a high Michaelis-Menten constant is useful for enhancing the  sensitivity to concentration modulation symbols and can lead to lower BER. 
}

\ifarxiv 

\fi

\appendices

\ifarxiv 
\section{Derivation of \eqref{eqn:ode:L}}
\label{app:sol:log_post_prob}
Recalling that ${\cal X}_*(t)$ is the history of \ce{X_*} in the time interval $[0,t]$. In order to derive (\ref{eqn:ode:L}), we consider the history ${\cal X}_*(t+\Delta t)$ as a concatenation of ${\cal X}_*(t)$ and ${\cal X}_*(\tau)$ for $\tau \in (t,t+\Delta t]$. We assume that $\Delta t$ is chosen small enough so that no more than one reaction or diffusion event can take place in $(t, t+\Delta t]$. Given this assumption and right continuity of continuous-time Markov chains, we can use $X_*(t+\Delta t)$ to denote the history of $X_*(t)$ in $(t,t+\Delta t]$. 

Consider the likelihood of observing the history ${\cal X}_*(t+\Delta t)$ assuming that the transmitter has sent symbol $s$: 
\begin{eqnarray}
&     & {\rm P}[{\cal X}_*(t+\Delta t)| s]  \label{eqn:star:like:1} \\
& = & {\rm P}[{\cal X}_*(t) \; \mbox{\textsc{and}} \; X_*(t+\Delta t) | s] \label{eqn:star:like:2} \\
& = & {\rm P}[{\cal X}_*(t) | {\cal H}_i]  \; {\rm P}[ X_*(t+\Delta t) | s, {\cal X}_*(t)] \label{eqn:star:like:3} 
\end{eqnarray}
where we have expanded ${\cal X}_*(t+\Delta t)$ in \eqref{eqn:star:like:1} using concatenation.

By using \eqref{eqn:star:like:3} in the definition of log-probability ratio in \eqref{eqn:def:L}, we can show that: 
\begin{align} 
L(t+\Delta t) = L(t) + \log \left( \frac{{\rm P}[X_*(t+\Delta t) | 1, {\cal X}_*(t)]}{{\rm P}[X_*(t+\Delta t) | 0, {\cal X}_*(t)]} \right) \label{eq:app:L}
\end{align}

The conditional probability ${\rm P}[X_*(t+\Delta t) | s, {\cal X}_*(t)]$ is the prediction of the number of \ce{X_*} molecules at time $t+\Delta t$ based on its history up till time $t$. This conditional probability can be obtained by solving an optimal Bayesian filtering problem over the continuous-time Markov chain or RDME that describes the dynamics of the molecular network. 
We considered how this conditional probability could be evaluated in our earlier work \cite{Awan:2017fm}. The key result in \cite{Awan:2017fm} says that ${\rm P}[X_*(t+\Delta t) | s, {\cal X}_*(t)]$ depends on the predicted rate of the chemical reactions in which \ce{X_*} are involved. By using \cite{Awan:2017fm}, we have: 
\begin{align}
&  {\rm P}[X_*(t+\Delta t) | s, {\cal X}_*(t)] =\nonumber \\ 
& \delta_{X_*(t+\Delta t),X_*(t) + 1} \; k_0 J_i(t_-) \; \Delta t +  \nonumber \\ 
& \delta_{X_*(t+\Delta t),X_*(t) - 1} \; g_- X_*(t) \; \Delta t \; + \nonumber  \\
& \delta_{X_*(t+\Delta t),X_*(t)}  \times \nonumber \\ 
& (1 - k_0 J_i(t) \; \Delta t - g_- X_*(t)  \; \Delta t) \label{eq:app:predictb}
\end{align}
where $\delta_{a,b}$ is the Kronecker delta which is 1 when $a$ equals to $b$ and zero otherwise, and  $J_i(t) =  {\rm E}[\{XK\}(t) | s, {\cal X}_*(t)]$ is the expected number of \cee{XK} molecules at time $t$ assuming that the transmitter has sent symbol $s$ and the history ${\cal X}_*(t)$. 

Note that ${\rm P}[X_*(t+\Delta t) | s, {\cal X}_*(t)]$ in \eqref{eq:app:predictb} is a sum of three terms with multipliers $\delta_{X_*(t+\Delta t),X_*(t) + 1}$, $\delta_{X_*(t+\Delta t),X_*(t) - 1}$ and $\delta_{X_*(t+\Delta t),X_*(t)}$. Since these multipliers are mutually exclusive, we have:
\begin{align}
 & \log \left( \frac{{\rm P}[X_*(t+\Delta t) | 1, {\cal X}_*(t)]}{{\rm P}[X_*(t+\Delta t) | 0, {\cal X}_*(t)]} \right)   \nonumber  \\
 = & 
 \delta_{X_*(t+\Delta t),X_*(t) + 1} \log \left( \frac{k_0 J_1(t_-) \; \Delta t  }{k_0 J_0(t_-) \; \Delta t }  \right) + \nonumber \\
 & \delta_{X_*(t+\Delta t),X_*(t) - 1} \log \left( \frac{g_- X_*(t) \; \Delta t}{g_- X_*(t) \; \Delta t}   \right) \nonumber + \\
 & \delta_{X_*(t+\Delta t),X_*(t)} \times \nonumber \\
& \log \left( \frac{ 1 - k_0 J_1(t_-) \; \Delta t - g_-  X_*(t)  \; \Delta t}{ 1 - k_0 J_0(t_-)  \; \Delta t - g_- X_*(t)  \; \Delta t }   \right) \nonumber \\
 \approx & 
\delta_{X_*(t+\Delta t),X_*(t) + 1} \log \left( \frac{J_1(t_-) }{J_0(t_-) }  \right) - \nonumber \\
& \delta_{X_*(t+\Delta t),X_*(t)} \; k_0  
\left( J_1(t) - J_0(t)  \right) \; \Delta t \label{eqn:app:likelihood} 
\end{align}
where we have used the approximation $\log(1 + f \; \Delta t) \approx f \; \Delta t$ and have ignored terms of order $(\Delta t)^2$ or higher to obtain \eqref{eqn:app:likelihood}. Note also that the above derivation assumes that $\frac{J_1(t)}{J_0(t)}$ is strictly positive so its logarithm is well defined; this can be achieved by proper choice of the hypotheses. 

By substituting \eqref{eqn:app:likelihood} into (\ref{eq:app:L}), we have after some manipulations and after taking the limit $\Delta t \rightarrow 0$:
\begin{align}
\frac{dL(t)}{dt} 
= & \lim_{\Delta t \rightarrow 0} \frac{\delta_{X_*(t+\Delta t),X_*(t) + 1} }{\Delta t}
\log \left( \frac{J_1(t_-)}{J_0(t_-)}  \right) - \nonumber \\
& \delta_{X_*(t+\Delta t),X_*(t)} \; k_0   
\left( J_1(t) - J_0(t)  \right) \label{eqn:app:logmapm1}  
\end{align}
In order to obtain (\ref{eqn:ode:L}), we use the following reasonings. First, the term $\lim_{\Delta t \rightarrow 0} \frac{\delta_{X_*(t+\Delta t),X_*(t) + 1} }{\Delta t}$ is a Dirac delta at the time instant that an \cee{X_*} molecule is produced. Since the instant is also the time at which $X_*(t)$ jumps by $+1$, we can identify this term with $\left[ \frac{ dX_*(t) }{dt} \right]_+$ where $[w]_+ = \max (w,0)$. 

Since $X_*(t)$ is a piecewise constant signal counting the number of \cee{X_*} molecules, its derivative is a sequence of Dirac deltas at the time instants that \cee{X} is activated or \cee{X_*} is deactivated. Note that the Dirac deltas corresponding to the activation of \cee{X} carries a positive sign and the $[  \; ]_+$ operator keeps only these. 
Fig.~\ref{fig:xstar_deri} shows an example of $X_*(t)$ and its corresponding $\left[ \frac{ dX_*(t) }{dt} \right]_+$.

\begin{figure}[t]
        \centering
        \includegraphics[scale = \picscale]{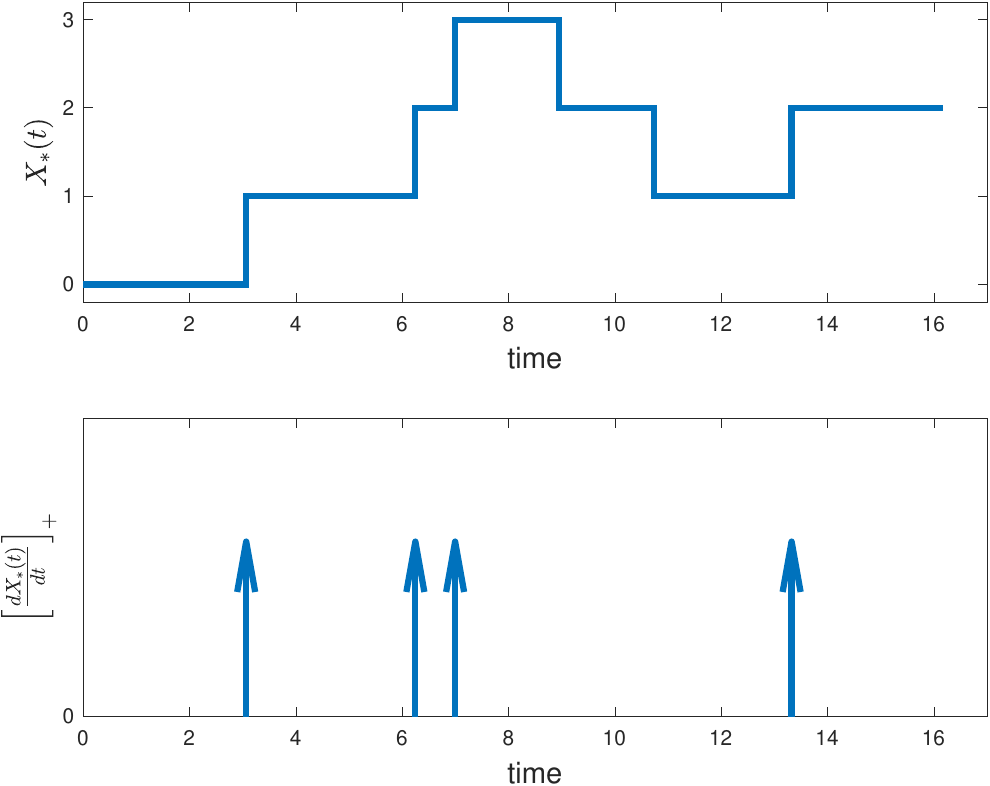} 
        \caption{The upper plot shows a sample $X_*(t)$ and the lower plot shows the corresponding $\left[ \frac{ dX_*(t) }{dt} \right]_+$.}
        \label{fig:xstar_deri}
\end{figure}

Second, the term $\delta_{X_*(t+\Delta t),X_*(t)}$ is only zero when the number of \ce{X_*} molecule changes but the number of such changes is countable. In other words, $\delta_{X_*(t+\Delta t),X_*(t)}=1$ with probability one. This allows us to drop $\delta_{X_*(t+\Delta t),X_*(t)}$. Hence \eqref{eqn:ode:L}.  
\fi

\section{Derivation for Sec.~\ref{sec:pdp:approx:front_end:para}}
\label{app:front_end}
\rev{
In this appendix, we carry out a steady-state analysis of the front-end circuit assuming that $K(t)$ is given by it steady state value of $[K]_s^{\rm ss}$. The reaction rate equations for the front-end species are: 
\begin{eqnarray} 
\frac{d[XK]}{dt} & = & 
a_0 [K] \; [X] - (d_0 + g_0) [XK] \\
\frac{d[X_*]}{dt} & = & 
g_0 [XK] - g_- [X_*]. 
\end{eqnarray} 
where we have dropped the dependence on $(t)$ to simplify the notation. By setting the derivatives to zeros and by using the conservation relation $[X] + [XK] + [X_*] = [X]_T$, we arrive at \eqref{eqn:Xstar:perturbation}. Another result from this steady state analysis is the steady state concentration of \ce{XK}, which is given by:
\begin{eqnarray}
[XK](t) & = & [X]_T
\frac{[K](t)}{H_{M0} + (1 + \gamma) [K](t)}
\label{eqn:XK:perturbation} 
\end{eqnarray} 
}

Let $[K]_s^{\rm ss}$ be the steady state concentration of \ce{K} when the transmitter sends Symbol $s$. 



\section{Approximate Bayesian filtering}
\label{app:filtering} 
The aim of this appendix is to derive an approximation of  $J_s(t) = {\rm E}[\{XK\}(t) | s, {\cal X}_*(t)]$ whose interpretation is the posteriori mean of the number of \cee{XK} molecules at time $t$ given the hypothesis that Symbol $s$ has been sent and the history ${\cal X}_*(t)$. We will derive an approximate expression for $J_s(t)$ as a function of the hypothesis $s$ and observations $X_*(t)$. Our derivation has 2 steps. The fist step uses the product Poisson entropic matching method in \cite{Bronstein:2018eh}. In the second step, we determine a closed-form approximation of the result in the first step. 

\rev{In order to simplify the presentation here, we will consider a medium consisting of 3 voxels arranged in a line. We will refer to these voxels as Voxels 0, 1 and Voxel 2.} (Generalisation to the more general case is straightforward and will be explained later.) We assume that the  transmitter and the receiver are located in, respectively, Voxels 0 and 2. 
\rev{
The transmitter injects vesicles into Voxel 1 at a rate of $r_{TX}$.
}  
\rev{
As in RDME, the movement of the molecules between the voxels is modelled by a unimolecular reaction. We assume that vesicles are lost with a rate constant of $\widetilde{d}_e$ and the rate constant for a vesicle to enter the receiver voxel is $\widetilde{d}_i$. Lastly, the signalling molecules degrades with a rate constant of $\widetilde{d}_d$ in the receiver voxel. 
} 

We first show how we can map our Bayesian filtering problem to the one considered in \cite{Bronstein:2018eh}, which is based on reaction counts. Let ${\cal R}_f(t)$ (resp. ${\cal R}_r(t)$) be the cumulative number of times that reaction \eqref{cr:p:z0:2} (resp. \eqref{cr:p:z0:3}) has taken place in the time interval $[0,t]$, i.e. ${\cal R}_f(t)$ and ${\cal R}_r(t)$ are time trajectories of reaction counts. We consider the Bayesian filtering problem which uses ${\cal R}_f(t)$ and ${\cal R}_r(t)$ as the given information. We argue that we can deduce ${\cal R}_f(t)$ and ${\cal R}_r(t)$ from ${\cal X}_*(t)$. This is because there is only one reaction in which \ce{X_*} is formed (i.e. reaction \eqref{cr:p:z0:2}) and only one reaction in which \ce{X_*} is deactivated (i.e. reaction \eqref{cr:p:z0:3}); therefore for a given ${\cal X}_*(t)$, the corresponding $R_f(t)$ and $R_r(t)$ can be uniquely determined. Overall, this implies that we can apply the approximation method in \cite{Bronstein:2018eh} to approximate $J_s(t)$. 

The Bayesian filtering problem is to use the observations ${\cal X}_*(t)$ and assumption $s$ to compute the posteriori probabilities of all possible system states that are compatible with the observations. Based on the 3-voxel medium set up which we assume in this appendix, the system state $v(t)$ is the 3-tuple $(K_1(t),K_2(t),\{XK\}(t))$ where $K_i(t)$ is the number of signalling molecules in Voxel $i$ at time $t$. The idea in \cite{Bronstein:2018eh} is to approximate the posteriori probability that the system is in a state by a product of independent Poisson distributions. Let $\theta_{s,1}(t)$, $\theta_{s,2}(t)$ and $\theta_{s,XK}(t)$ denote the means of three independent Poisson distributions in our problem set up. These three Poisson means are used to approximate the posteriori means, as follows: $\theta_{s,i}(t) \approx {\rm E}[K_i(t) | s, {\cal X}_*(t)]$ ($i = 1, 2$) and $\theta_{s,XK}(t) \approx {\rm E}[\{XK\} | s, {\cal X}_*(t)]$. In particular,  $\theta_{s,XK}(t)$ is the approximation that we are seeking. In the following, we will drop the $(t)$ dependence for brevity. 

We now present the result of applying the method in \cite{Bronstein:2018eh} to the 3-voxel set up. The result states how the approximate posteriori means $\theta_{s,1}$, $\theta_{s,2}$ and $\theta_{s,XK}$ evolve over time. The evolution of these means consist of both discrete jumps and continuous change. The posteriori mean $\theta_{s,XK}$ experiences a discrete jump at the time when an \ce{X_*} is formed; at other times, according to \cite[Eq.~(33)]{Bronstein:2018eh}, the evolution of $\theta_{s,i}$ and $\theta_{s,XK}$ obeys the following ODEs:
\rev{
\begin{subequations}
\begin{align}
\frac{d\theta_{s,1}}{dt} =& r_{TX} - (\widetilde{d}_e + \widetilde{d}_i) \theta_{s,1}
\label{app:theta_i:ode:1} \\
\frac{d\theta_{s,2}}{dt} =& -\widetilde{d}_d \theta_{s,2} + \widetilde{d}_i \theta_{s,1} - \frac{1}{\Omega} a_0 \theta_{s,2} \theta_{s,X} + d_0 \theta_{s,XK} 
\label{app:theta_i:ode:2} \\
\frac{d\theta_{s,XK}}{dt} =& 
 \frac{1}{\Omega} a_0 \theta_{s,2} \theta_{s,X} - (d_0+g_0) \theta_{s,XK} 
\label{app:theta_i:ode:SK}  
\end{align}
\label{app:theta_i:ode:all}  
\end{subequations}
}\noindent{where} $\theta_{s,X} = X_T - X_*(t) - \theta_{s,XK}$ and $X_T$ is the total number of substrate molecules in the receiver voxel. This completes the first step which is to apply the method of \cite{Bronstein:2018eh} to our problem. 

Our next step is to find an approximation for $\theta_{s,XK}$. Given the history ${\cal X}_*(t)$, let $t_1$, $t_2$, ... be time instants that ${\cal X}_*(t)$ experiences a jump because an \ce{X_*} molecule is produced or reverted. In each time interval $(t_j,t_{j+1})$, the number of \ce{X_*} molecules is a constant and we denote that by $X_{*,j}$. We note that $\theta_{s,XK}$ in \eqref{app:theta_i:ode:SK} is a fast variable because $d_0, g_0 \gg a_0$ so $\theta_{s,XK}$ reaches steady state quickly. Our proposal is to approximate $\theta_{s,XK}$ by a constant value in each time interval $(t_j,t_{j+1})$ where the constant value is the steady state solution of \eqref{app:theta_i:ode:all} assuming that $X_*(t)$ equals to $X_*^{[j]}$ for all $t \in [0,\infty)$. Note that our approximation for $\theta_{s,XK}$ is a piecewise constant trajectory over time and the value of the $\theta_{s,XK}$ depends on the value of $X_*^{[j]}$ and the kinetic parameters. 

Our next step is to determine the steady state solution of \eqref{app:theta_i:ode:all} assuming that $X_*(t)$ equals to $X_*^{[j]}$ $\forall t \in [0,\infty)$. We will use $\tilde{\theta}_{s,i}$ and $\tilde{\theta}_{s,XK}$ to denote the steady state solution of \eqref{app:theta_i:ode:all}. By setting the LHSs of \eqref{app:theta_i:ode:all} to zero, we have, after some manipulations:
\rev{
\begin{align}
\tilde{\theta}_{s,i} =& \alpha_{RX,TX} r_{\rm TX} - g_0 \tilde{\theta}_{s,SK} 
\label{eqn:app:baye:theta_s2} \\
\tilde{\theta}_{s,XK} =& \frac{1}{\Omega H_{M0}} \tilde{\theta}_{s,2} \; 
(X_T - X_*^{[j]} - \tilde{\theta}_{s,XK}) 
\label{eqn:app:baye:ss:theta_sk}
\end{align}
}

\rev{
where $\alpha_{RX,TX} = \frac{\tilde{d}_i}{\tilde{d}_e + \tilde{d}_i}$. The constant $-\alpha_{RX,TX}$ quantifies the transfer of molecules from the transmitter voxel to the receiver voxel. We can identify $\alpha_{RX,TX} r_{\rm TX}$ as the steady state mean number of \ce{K} molecules in the receiver voxel, so we will denote it by $K_s^{\rm ss}$ which is the notation that we have used in the main text to denote such as quantity.} 

By substituting the expression of $\tilde{\theta}_{s,2}$ in \eqref{eqn:app:baye:theta_s2} into the RHS of \eqref{eqn:app:baye:ss:theta_sk}, we can show that $\tilde{\theta}_{s,XK}$ is the smaller root of the quadratic equation $q_2 x^2 + q_1 x + q_0 = 0$ in the indeterminate $x$ where the coefficients are given by: 
\rev{
\begin{align}
q_2 =& \frac{g_0}{\Omega} \\
q_1 =& -(g_0 \frac{X_T - X_*^{[j]}}{\Omega} +  \frac{K_{s}^{\rm ss}}{\Omega} + H_{M0})  \\
q_0 =& \frac{K_{s}^{\rm ss}}{\Omega} (X_T - X_*^{[j]}) 
\end{align}
}
\rev{
Under the condition that $H_{M0} \gg \min(g_0 \frac{X_T}{\Omega}, \frac{K_{s}^{\rm ss}}{\Omega})$, we can apply the approximation in \cite{Straube.2017sal} to show that $\tilde{\theta}_{s,XK} \approx -\frac{q_0}{q_1}$ or
\begin{align}
\tilde{\theta}_{s,XK} \approx 
\frac{\frac{K_{s}^{\rm ss}}{\Omega} (X_T - X_*^{[j]})}{g_0 \frac{X_T - X_*^{[j]}}{\Omega} +  \frac{K_{s}^{\rm ss}}{\Omega} + H_{M0}}
\label{eqn:app:baye:theta_SK:one_interval}
\end{align}
}
Since the term $H_{M0}$ in the denominator is much greater than the other two terms, we will replace $X_*^{[j]}$ by the steady state mean of $X_*(t)$ assuming Symbol $s$ is sent, which we will denote by $\bar{X}_{s}$. This is so that the denominator is independent of the observation $X_*^{[j]}$. Furthermore, the RHS of \eqref{eqn:app:baye:theta_SK:one_interval} is the expression for the time interval $(t_j, t_{j+1})$. Since $X_*^{[j]} = X_*(t) \forall t \in (t_j, t_{j+1})$, we can therefore obtain a $\tilde{\theta}_{s,XK}$ which holds for all $t$ by replacing $X_*^{[j]}$ in the RHS of \eqref{eqn:app:baye:theta_SK:one_interval} by $X_*(t)$. After that, we obtain \eqref{eqn:Js:approx} where we have used $\tilde{\theta}_{s,XK}$ as the approximation of $J_s(t)$. 

Finally, we consider the condition 
\rev{
\begin{align}
H_{M0} \gg \min(g_0 \frac{X_T}{\Omega}, \frac{K_{s}^{\rm ss}}{\Omega})  
\label{eqn:app:baye:approx:condition}
\end{align}
}
for the approximation in this Appendix to hold. The quantity $\frac{K_{s}^{\rm ss}}{\Omega}$ is the concentration of the signalling molecules in the receiver voxel when the receiver is absent. We consider this quantity in Sec.~\ref{sec:pdp:approx:front_end:para} and denote it by $\bar{[K]}_{s,RX}$. In Sec.~\ref{sec:pdp:approx:front_end:para}, we impose the condition $H_{M0} \gg \bar{[K]}_{s,RX}$ so that the front-end molecular circuit has high \rev{Michaelis-Menten constant}. It can be shown that if $H_{M0} \gg \bar{[K]}_{s,RX}$ holds, then \eqref{eqn:app:baye:approx:condition} holds. Therefore, the condition needed for high \rev{Michaelis-Menten constant} is sufficient for the results in this Appendix to hold. 


\rev{The derivation above assumes there are only 3 voxels and the voxels are arranged in a specific way. However, it can be shown that \eqref{eqn:app:baye:theta_SK:one_interval} holds in general. 
}

\ifarxiv 
\section{Derivation for \eqref{eqn:pdp:llr:Xstar_only} }   	
\label{app:Lhat}

The aim of this appendix is to show that the log-probability ratio computation in \eqref{eqn:ode:L:kappa} can be approximated by  \eqref{eqn:pdp:llr:Xstar_only}. We begin by writing \eqref{eqn:ode:L:kappa} in the integral form:
\begin{align}
L(T) 
= & \int_0^T \left[ \frac{ dX_*(t) }{dt} \right]_+  
\log \left( \frac{\kappa_1}{\kappa_0} \right) dt \; - 
\nonumber \\
& k_0  \left( \kappa_1 - \kappa_0 \right) \int_0^T (X_T - X_*(t)) dt
\label{eqn:app:L:int:kappa}  
\end{align}

The derivation is divided into 2 steps where each step focuses on deriving an approximation for one of the integrals in \eqref{eqn:app:L:int:kappa}.  \\

\noindent{}\textbf{Step 1: Approximating the first integral in \eqref{eqn:app:L:int:kappa}}\\
The first integral on the RHS of \eqref{eqn:app:L:int:kappa} can be interpreted as $T$ times the mean production rate of \ce{X_*} molecules. 
Over a large $T$, we have:
\begin{align}
\lim_{T \rightarrow \infty} \frac{1}{T} 
\int_0^T \left[ \frac{ dX_*(t) }{dt} \right]_+ dt 
 = &  
\lim_{T \rightarrow \infty} \frac{1}{T} 
g_0 \int_0^T  \{XK\}(t) ) dt  
\label{eqn:app:timeavg:Xstar:prod} 
\end{align}
where the RHS of the above equation models the production of \ce{X_*} from \ce{XK} according to Reaction \ref{cr:p:z0:2}. Next, at steady state, the production rate of \ce{X_*} is balanced by its reversing rate, hence we have: 
\begin{align}
\lim_{T \rightarrow \infty} \frac{1}{T} 
g_0 \int_0^T  \{XK\}(t) ) dt 
 = &  
\lim_{T \rightarrow \infty} \frac{1}{T} 
\int_0^T g_- X_*(t)  dt 
\label{eqn:app:timeavg:Xstar:prod:revert} 
\end{align}
By combining \eqref{eqn:app:timeavg:Xstar:prod} and \eqref{eqn:app:timeavg:Xstar:prod:revert}, we can therefore approximate the first integral on the RHS of \eqref{eqn:app:L:int:kappa} by:
\begin{align}
\int_0^T \left[ \frac{ dX_*(t) }{dt} \right]_+ dt 
 \approx &  
\int_0^T g_- X_*(t)  dt
\label{eqn:app:timeavg:Xstar:diffLapprox} 
\end{align}

\noindent{}\textbf{Step 2: Approximating the second integral in \eqref{eqn:app:L:int:kappa}} \\
We now move onto the second integral on the RHS of \eqref{eqn:app:L:int:kappa}. The derivation here assumes that the system is in steady state, this allows us to replace the time average in by its ensemble average. In this part, we will overload the symbols $X$, $X_*$ and $\{XK\}$ to refer to the random variables of the number of, respectively, \ce{X}, \ce{X_*} and \ce{XK} molecules at steady state. This should not cause any confusion because the meaning should be clear from the context. With this overloading, the mean number of \cee{X} at steady state is denoted by ${\rm E}[X]$ etc. We first need to state or derive a number of auxiliary results. \\

We first need make a clarification on the notation. The reason we add a pair of curly brackets to the notation $\{XK\}$, which denotes the random variable of the number of \ce{XK} molecules in steady state, is to stress that it is referring to complex \ce{XK}. This is important because we will also be multiplying the random variables $X$ and $K$ in the derivation and we will write the multiplication as $X \cdot K$. \\

We will now derive or present a number of auxiliary results. After that we will combine all these auxiliary results to arrive at an approximation for the second integral in \eqref{eqn:app:L:int:kappa}. \\

\noindent{\sl (Auxiliary Result 1)} At steady state, the production and reversion of the \ce{XK} molecules balance out, therefore, we have:
\begin{align}
\frac{a_0}{\Omega} {\rm E}[X \cdot K] = & 
(d_0 + g_0) {\rm E}[ \{XK\} ]
\label{eqn:app:ensemble:XK} 
\end{align}

In terms of ensemble averages, we can rewrite \eqref{eqn:app:timeavg:Xstar:prod}  as $ g_0 {\rm E}[ \{XK\} ] = g_- {\rm E}[X_*]$. By combining this with \eqref{eqn:app:ensemble:XK}, we have Auxiliary Result 1:
\begin{align}
{\rm E}[X \cdot K] = & 
\frac{H_{M0} \Omega}{\gamma}  {\rm E}[ X_* ]
\label{eqn:app:Lhat:aux:1} 
\end{align}

\noindent{\sl (Auxiliary Result 2)} Since the receiver front-end circuit has high \rev{Michaelis-Menten constant}, it means that we can approximate the signalling molecule count in the receiver voxel when the front-end is present by the one when the receiver is absent, we will therefore ignore the present of the front-end for deriving this auxiliary result. Since the diffusion of the molecules is independent, the distribution of the signalling molecules in the receiver voxel is binomial. We will state a property of the approximation of the reciprocal of the mean of a binomial variable in this auxiliary result. \\ 

Consider a binomial distribution $B(Q; m, f)$ with parameters $m$ (number of trials) and $f$ (success probability), then for sufficiently large $m$ and $f$, we have 
\begin{align}
\frac{1}{{\rm E}[Q]} \approx& \; {\rm E}[I(\frac{1}{Q})] \label{app:L2:aux:3}
\end{align}
where 
\begin{align}
I(\frac{1}{q}) =&
\left\{
\begin{array}{cl}
0               & \mbox{for } q = 0          \\
\frac{1}{q} & \mbox{for } q \geq 1
\end{array} 
\right.
\label{eqn:app:Lhat:aux:2} 
\end{align}
This result essentially says that the mean of the reciprocal of a binomial random variable (with $\frac{1}{0}$ excluded) is approximately equal to the reciprocal of the mean of the binomial random variable. If $f = 1$ and $m \geq 1$, the binomial distribution has a single outcome with a non-zero probability so \eqref{eqn:app:Lhat:aux:2}  is exact. Intuitively, if a probability has a single modal distribution with a narrow spread, then \eqref{eqn:app:Lhat:aux:2} holds approximately. For $f = 0.1$, the relative error of using \eqref{eqn:app:Lhat:aux:2}  is 3.21\% for $m = 300$ and drops to 1.87\% for $m = 500$. 
In general, the approximation is better for large $m$ and $f$. \\

\noindent{\sl (Auxiliary Result 3)} Since the forward reaction in \eqref{cr:p:z0:1} is slow because $a_0$ is assumed to be small, we can show that:
\begin{align}
{\rm E}[(X_T - X_*) \cdot K] \approx& \; {\rm E}[X_T - X_*]    \; {\rm E}[K]   \label{eqn:app:Lhat:aux:3_1} 
\end{align}

We first write ${\rm E}[ X \cdot K]$ as a time integral:
\begin{align}
{\rm E}[X \cdot K] = \int_{t = 0}^\infty X(t) K(t) dt 
\label{eqn:app:Lhat:aux:3_2}
\end{align} 
The reason why we are considering this integral is because \ce{X} and \ce{K} are the reactants of the slow reaction in \eqref{cr:p:z0:1}. This means that the counts of \ce{X} change slower than that of \ce{K}. 
This difference in time-scale allows us to approximate the above integral.  \\

Let $t_0$, $t_1$, $\ldots$ be a sequence of time instants at which $X(t)$ changes its value. We will re-write the integral on the RHS of \eqref{eqn:app:Lhat:aux:3_2} as a sum of integrals: 
\begin{align}
\int_{t = 0}^\infty X(t) K(t) dt =&  
\sum_{i = 0}^{\infty}  X(t_i) \int_{t_i}^{t_{i+1}} K(t) \; dt 
\label{eqn:app:Lhat:aux:3_3} 
\end{align}

Since $X(t)$ is slow in comparison with $K(t)$, this means the time interval $[t_i,t_{i+1})$ is likely to be long compared to the time-scale of the faster $K(t)$. This allows us to approximate the integral on the RHS of \eqref{eqn:app:Lhat:aux:3_3} by ${\rm E}[K] (t_{i+1} - t_i)$. Hence we have:
\begin{align}
{\rm E}[X \cdot K] =& \sum_{i = 0}^{\infty}  X(t_i) E[K] \; dt \nonumber \\
=& {\rm E}[X] \; {\rm E}[K] 
\end{align}
Note that the above argument is identical to the one used in \cite{Cao:2005gj} to derive the slow-scale tau-leaping simulation algorithm. \\

Next, we do two substitutions. First, we note that Auxiliary Result 1 expresses ${\rm E}[X \cdot K]$ in terms of ${\rm E}[X_*]$. Second, we approximate ${\rm E}[X]$ by ${\rm E}[X_T - X_*]$ because we know the front-end circuit has high \rev{Michaelis-Menten constant} so ${\rm E}[\{ XK \}]$ is small. After these substitutions, we arrive at Auxiliary Result 3:
\begin{align}
{\rm E}[X_T - X_*]  \approx \;  \frac{H_{M0} \Omega}{\gamma} 
\frac{{\rm E}[E_*]}{ {\rm E}[K]}
\end{align}

\noindent{\sl (Auxiliary Result 4)} By using the same argument as in Auxiliary Result 3, we can show that:
\begin{align}
{\rm E}[X_* I(\frac{1}{K})] \approx& \; {\rm E}[X_*]    \; {\rm E}[I(\frac{1}{K})]   \label{app:L2:aux:3_2} 
\end{align}
\\

By using the above auxiliary results we have
\begin{align}
& {\rm E}[X_T - X_*]  & \nonumber \\
\approx&  \frac{H_{M0} \Omega}{\gamma} 
\frac{{\rm E}[E_*]}{ {\rm E}[K]} \; \; &\mbox{(Aux. Result 3)} \nonumber \\
=& \frac{H_{M0} \Omega}{\gamma} {\rm E}[E_*] {\rm E}[I(\frac{1}{K})]
\nonumber \; \; &\mbox{(Aux. Result 2)}  \\
=& \frac{H_{M0} \Omega}{\gamma} {\rm E}[X_* I(\frac{1}{K})] \; \; &\mbox{(Aux. Result 4)}
\end{align}

By applying the results in the 2 steps above, we can show that \eqref{eqn:ode:L:kappa} can be approximated by:
\begin{align}
\frac{L(t)}{dt} = & 
g_- X_*(t) 
\left[
\log \left( \frac{\kappa_1}{\kappa_0} \right) - H_{M0} \Omega \frac{ \kappa_1 - \kappa_0  }{K(t)}   
\right]
\label{eqn:app:pdp:llr_approx}
\end{align}
which is \eqref{eqn:pdp:llr:Xstar_only}. 

\fi

\ifarxiv
\section{Deriving \eqref{eqn:pdp:hyperbolic}} 
\label{app:hyperbolic} 
In this Appendix, we will show that the enzymatic cycle \eqref{cr:p:Z1:all}, which is referred to as the threshold-hyperbolic (TH) cycle in Sec.~\ref{sec:Lhat:TH}, can be used to realise a TH function. Since the TH-cycle has multiple non-linearities, a stochastic analysis is not tractable. Instead, we will use quasi-steady state analysis \cite{Gomez-Uribe:PLoS_CB:2005} which is also used in Appendix \ref{app:front_end}. We further simplify the analysis by not including the diffusion of \ce{K} in the model. We justify this simplification by the fact that we will require the TH-cycle to have high \rev{Michaelis-Menten constant} which means few \ce{K} molecules will be sequestered by this cycle. This in turn means that the behaviour of the TH-cycle can be analyse without considering the details on diffusion. Based on these simplifications, we assume that the total count of \ce{K} molecules seen by the cycle is a constant and we will denote this by $K_T$ where $K_T$ can be considered to be the mean steady state count of signalling molecules in the receiver voxel. We expect $K_T = [K]_{s}^{\rm ss} \Omega$. \\

We will show that under the assumptions that $H_{M1} \gg \frac{K_T}{\Omega}$, $H_{M2} \ll \frac{P_T}{\Omega}$ and $\frac{k_1 K_T}{k_2 P_T}$ is sufficiently large, then the amount of \ce{Y_*} in the TH-cycle is a threshold-hyperbolic function of $K_T$ which is the input level of the cycle. \\

Our derivation makes use of the results in \cite{Straube.2017sal} which uses quasi-steady analysis to study the properties of enzymatic cycles of the form \eqref{cr:p:Z1:all}. Since the quasi-steady state analysis uses concentration rather than counts, we will temporarily switch over to concentration so that the reader can better match the formulas here with those in \cite{Straube.2017sal}. Note that all the concentrations in this Appendix are steady state concentrations. \\

The implication of the assumption $H_{M2} \ll [P]_T$ has been studied in \cite{Straube.2017sal}. Let ${\cal Y} = [Y_*] + [Y_*P]$. By using  \cite[Eq.~(29)]{Straube.2017sal}, which holds when $H_{M2} \ll [P]_T$, we have: 

\begin{eqnarray}
[Y_*P] & \approx & 
\left\{
\begin{array}{ll}
{\cal Y}  \left(  1 - \frac{ H_{M2}    }{  [P]_T - {\cal Y} }  \right)  &  \mbox{ for }  {\cal Y} < [P]_T \\
\hspace{0mm} [P]_T  \left(  1 - \frac{ H_{M2}    }{ {\cal Y}  - [P]_T}  \right) &  \mbox{ for }  {\cal Y} > [P]_T 
\end{array}
\right.
\end{eqnarray}

Since we assume $H_{M2}$ is small, we set $H_{M2}$ to zero in the above expressions. After some simplification, we have:
\begin{align}
[Y_*] & \approx 0          &\mbox{ for }  [Y_*] + [Y_*P] < [P]_T 
\label{eqn:app:Z1:Ystar:ss}
\\
\hspace{0mm} 
[Y_*P] & \approx [P]_T   &\mbox{ for }  [Y_*] + [Y_*P] > [P]_T 
\label{eqn:app:Z1:YstarP:ss}
\end{align}

Note that $[Y_*] + [Y_*P]$ is small when the input level is small, and vice versa. This derivation shows that if the input level is low, then according to \eqref{eqn:app:Z1:Ystar:ss} we have $[Y_*] \approx 0$ or few 
\ce{Y_*}. This is the threshold part of the TH function. On the other hand, if the input level is high, then according to \eqref{eqn:app:Z1:YstarP:ss}, we have $[Y_*P] \approx  [P]_T$ or \ce{P} is saturated. \\

We now focus on deriving an expression for $[Y_*]$ when the input is high. The assumption $H_{M1} \gg \left( 1 + \frac{k_1}{k_2} \right) K_T$ implies that the following holds \cite[Eq.~(67)]{Straube.2017sal}: 
\begin{align}
& [Y_*] + [Y_*P]  \nonumber \\ = & [Y]_T - \frac{[K]_T + H_{M1}}{\alpha-1} \mbox{ for }  [Y_*] + [Y_*P] > [P]_T 
\end{align}
where $\alpha = \frac{k_1 [K]_T}{k_2 [P]_T}$. By suitable choice of the enzymatic cycle parameters, we can make $\alpha > 1$ and recall that $[Y_*P] \approx  [P]_T$, therefore we have:
\begin{eqnarray}
[Y_*]   & = & \widetilde{h}_0 - \frac{\widetilde{h}_1}{[K]_T} 
\label{eqn:app:threshyper:Y*}
\end{eqnarray}
where $\widetilde{h}_0 = [Y]_T - [P]_T - \frac{k_2}{k_1} [P]_T$ and $\widetilde{h}_1 = \frac{k_2}{k_1} [P]_T H_{M1}$. This shows that the concentration of \ce{Y_*} (i.e., $[Y_*]$) is a hyperbolic function of the input level $[K]_T$ when the input is high. The result \eqref{eqn:app:threshyper:Y*} is expressed in concentration, by multiplying its both sides by $\Omega$, we obtain \eqref{eqn:pdp:hyperbolic}. 
\fi

\ifarxiv
\section{Parameters for the TH-cycle}
\label{app:TH:para} 
This appendix explains how we determine the rate constants of the TH-cycle \eqref{cr:p:Z1:all} to realise the scaled TH function:
\begin{eqnarray}
\left[
\underbrace{
\rho \log \left( \frac{\kappa_1}{\kappa_0} \right)
}_{\zeta_0} 
\; - \;
\underbrace{
\rho H_{M0} (\kappa_1 - \kappa_0)  
}_{\zeta_1} 
\; \frac{\Omega}{K_T}  
\right]_+ 
\label{eqn:app:th:hyper}
\end{eqnarray}
where $\rho > 0$ is a scaling constant. The value of $\rho$ can be chosen so that the probability that \ce{Y} is active is high when the number of \ce{K} molecules is high. Note that if one scales the log-probability ratio \eqref{eqn:pdp:llr_approx} and the detection decision threshold by the same constant, the property of the detector does not change, therefore the scaling mentioned above is allowed. 

We assume that $\rho$ has been chosen beforehand, so we can assume $\zeta_0$ and $\zeta_1$ are given. In addition, as stated in Sec.~\ref{sec:Lhat:X_times_Y}, we use molecules of the form \ce{X-Y} to realise the computation of on the RHS of \eqref{eqn:pdp:llr_approx}. This means that $[X]_T = [Y]_T$. Here, we assume that the concentration $[X]_T$ has been chosen and that means the concentration $[Y]_T$ has been fixed. 

By comparing \eqref{eqn:app:th:hyper} and \eqref{eqn:app:threshyper:Y*}, we have:
\begin{eqnarray}
 [Y]_T - [P]_T - \frac{k_2}{k_1} [P]_T & = & \zeta_0 \\
 \frac{k_2}{k_1} [P]_T H_{M1} & = & \zeta_1 
\end{eqnarray}
The Michaelis-Menten constant $H_{M1}$ needs to be much bigger than $[K]_T$ for the cycle \eqref{cr:p:Z1:all} to behave as a threshold-hyperbolic function. We arbitrarily choose $H_{M1}$ to be 10 times larger than the maximum $[K]_T$ that the PH-cycle will encounter which happens when Symbol 1 is sent. Once $H_{M1}$ has been fixed, the values of $\frac{k_1}{k_2}$ and $[P]_T$ can be solved from the above two equations for the given $[Y]_T$. \\

Since we need $[P]_T \gg H_{M2}$ for the threshold-hyperbolic behaviour, we set $H_{M2} = \frac{[P]_T}{80}$ where 80 is an arbitrary choice. As mentioned in Sec.~\ref{sec:Lhat:X_times_Y}, we choose $k_1$ and $k_2$ so that the reaction time-scales of \ce{X} and \ce{Y} are similar. However, note that we have determined the ratio $\frac{k_1}{k_2}$ earlier so the choice of $k_1$ and $k_2$ need to satisfy this ratio. Finally, for the other reaction rate constants, we make an arbitrary choice of $d_1 = k_1$ and $d_2 = k_2$, then $a_1$ and $a_2$ are computed from the definitions of $H_{M1}$ and $H_{M2}$. \\
\fi

\ifarxiv
\section{RHS of \eqref{eqn:pdp:llr_approx} is proportional to the number \ce{X_*-Y_*} molecules} 
\label{app:Xp_times_Yp} 

The aim of this appendix is to show that, at steady state, the RHS of \eqref{eqn:pdp:llr_approx} is proportional to the number \ce{X_*-Y_*} molecules. An assumption that we need in our proof is that the activations of \ce{X} and \ce{Y} by \ce{K} are independent, i.e., we have: 
\begin{align}
& {\rm P}[ \mbox{\ce{X-Y} in state \ce{X_*-Y_*}} ] \nonumber \\  
= & 
{\rm P}[ \mbox{\ce{X} site in \ce{X-Y} is in  \ce{X_*} state} ] 
\times \nonumber \\
& 
{\rm P}[ \mbox{\ce{Y} site in \ce{X-Y} is in  \ce{Y_*} state} ]
\label{eqn:app:prob:Xp-Yp} 
\end{align} 
Let $\{X-Y\}_T$ be the total number of \ce{X-Y} molecules in its various states. Also, let $\{X_*-Y_*\}^{\rm ss}$ be the steady state number of molecules in the \ce{X_*-Y_*} state, $X_*^{\rm ss}$ (resp. $Y_*^{ss}$) is steady state number of molecules where the \ce{X} (\ce{Y}) site in the \ce{X_*} (\ce{Y_*}) state. We can rewrite \eqref{eqn:app:prob:Xp-Yp} as:
\begin{align}
\frac{\{X_*-Y_*\}^{\rm ss}}{\{X-Y\}_T} = 
\frac{X_*^{\rm ss}}{\{X-Y\}_T} \times 
\frac{Y_*^{\rm ss}}{\{X-Y\}_T}
\label{eqn:app:prob:Xp-Yp_propto_Xp_times_Yp} 
\end{align}

We now start with the RHS of \eqref{eqn:pdp:llr_approx} at steady state, which is proportional to the product of $X_*^{\rm ss}$ and that of the TH-function. Since we show in Appendix \ref{app:hyperbolic} that the TH-function is proportional to $Y_*^{ss}$, therefore, the RHS of \eqref{eqn:pdp:llr_approx} is proportional to the product $X_*^{\rm ss} Y_*^{\rm ss}$. By \eqref{eqn:app:prob:Xp-Yp_propto_Xp_times_Yp}, we can therefore conclude that the RHS of \eqref{eqn:pdp:llr_approx} at steady state is proportional to the number of molecules  in the \ce{X_*-Y_*} state. 
\fi

\section{Parameters of the transmitter and enzymatic cycles}
\label{app:cycle:para} 

For the transmitter, we assume that 20 and 120 mRNA molecules are used to produce, respectively, Symbols 0 and 1. The parameters of the cycle \eqref{cr:p:Z1:all} are: 
$a_1$ = 0.015 nM$^{-1}$s$^{-1}$;
$d_1$ = 5s$^{-1}$;
$k_1$ = 5s$^{-1}$;
$a_2$ = 4.8424nM$^{-1}$;
$d_2$ = 0.785s$^{-1}$;
$k_2$ = 0.785s$^{-1}$; and, 
$P_T$ = 16.6nM.  
The parameters of the cycle \eqref{cr:p:int:all} are: 
$a_3$ = 1.2$\times 10^{-4}$ nM$^{-1}$s$^{-1}$;
$d_3$ = 100s$^{-1}$;
$k_3$ = 100s$^{-1}$;
$a_4$ = 1.2$\times 10^{-6}$ nM$^{-1}$s$^{-1}$;
$d_4$ = 1s$^{-1}$;
$k_4$ = 1s$^{-1}$; 
$J_T$ = 308nM; and,
$\tilde{P}_T$ = 50nM.

\ifarxiv
\section{Diffusion limited binding rate for enzymatic cycles}
\label{app:diff_limited}
This appendix estimates the diffusion limited binding rate constant for the bimolecular reaction in the enzymatic cycle. Consider two generic molecules called \ce{A} and \ce{B}. We denote their diffusion coefficients as $D_A$ and $D_B$, and their radii $r_A$ and $r_B$. The diffusion limited binding rate constants for these molecules can be estimated by $4 \pi (D_A + D_B) (r_a + r_b)$ \cite{Takahashi:2010ko}. In our case, the bimolecular reactions are between a kinase and a substrate, thus both of them are protein. We use the upper limit of diffusion rate of 100 $\mu$m$^2$/s for protein \cite{Zhang_Hess:2019}. We take the sum $(r_a + r_b)$ to be 5nm \cite{Takahashi:2010ko}. This gives a diffusion limited binding rate of 7.5 nM$^{-1}$s$^{-1}$.

\fi


\end{document}